\documentclass[
superscriptaddress,
amsmath,
amssymb,
aps,
]{revtex4-2}
\usepackage{array}
\usepackage{multirow}
\usepackage{graphicx}
\usepackage{subfigure}
\usepackage{booktabs}
\usepackage{dcolumn}% Align table columns on decimal point
\usepackage{bm}% bold math
\usepackage{siunitx}
\usepackage{lineno}
\usepackage{xcolor}
\usepackage{comment}
\usepackage{caption}
\usepackage{adjustbox}
\usepackage[utf8x]{inputenc}
\usepackage{ucs}
\usepackage{ulem}
\usepackage{supertabular}
\usepackage{hyperref}% add hypertext capabilities
\usepackage{cleveref}

\begin{document}
\title{Neutron-induced nuclear recoil background in the PandaX-4T experiment}
% !TEX root = neutron_main.

\def\shKeyLab{School of Physics and Astronomy, Shanghai Jiao Tong University, Key Laboratory for Particle Astrophysics and Cosmology (MoE), Shanghai Key Laboratory for Particle Physics and Cosmology, Shanghai 200240, China}
\def\USTClab{State Key Laboratory of Particle Detection and Electronics, University of Science and Technology of China, Hefei 230026, China}
\def\USTCdep{Department of Modern Physics, University of Science and Technology of China, Hefei 230026, China}
\def\BUAA{School of Physics, Beihang University, Beijing 102206, China}
\def\BUAALab{Beijing Key Laboratory of Advanced Nuclear Materials and Physics, Beihang University, Beijing, 102206, China}
\def\zzu{School of Physics and Microelectronics, Zhengzhou University, Zhengzhou, Henan 450001, China}
\def\pku{School of Physics, Peking University, Beijing 100871, China}
\def\YaLongSD{Yalong River Hydropower Development Company, Ltd., 288 Shuanglin Road, Chengdu 610051, China}
\def\IAP{Shanghai Institute of Applied Physics, Chinese Academy of Sciences, 201800 Shanghai, China}
\def\CHEPpku{Center for High Energy Physics, Peking University, Beijing 100871, China}
\def\SDUdep{Research Center for Particle Science and Technology, Institute of Frontier and Interdisciplinary Scienc, Shandong University, Qingdao 266237, Shandong, China}
\def\SDUlab{Key Laboratory of Particle Physics and Particle Irradiation of Ministry of Education, Shandong University, Qingdao 266237, Shandong, China}
\def\UMD{Department of Physics, University of Maryland, College Park, Maryland 20742, USA}
\def\TDLee{Tsung-Dao Lee Institute, Shanghai Jiao Tong University, Shanghai, 200240, China}
\def\MESJTU{School of Mechanical Engineering, Shanghai Jiao Tong University, Shanghai 200240, China}
\def\SYU{School of Physics, Sun Yat-Sen University, Guangzhou 510275, China}
\def\NKU{School of Physics, Nankai University, Tianjin 300071, China}
\def\FDU{Key Laboratory of Nuclear Physics and Ion-beam Application (MOE), Institute of Modern Physics, Fudan University, Shanghai 200433, China}
\def\USST{School of Medical Instrument and Food Engineering, University of Shanghai for Science and Technology, Shanghai 200093, China}
\def\SJTUSC{Shanghai Jiao Tong University Sichuan Research Institute, Chengdu 610213, China}
\def\Princeton{Physics Department, Princeton University, Princeton, NJ 08544, USA}
\def\MIT{Department of Physics, Massachusetts Institute of Technology, Cambridge, MA 02139, USA}
\def\SARI{Shanghai Advanced Research Institute, Chinese Academy of Sciences, Shanghai 201210, China}
\def\SPEIT{SJTU Paris Elite Institute of Technology, Shanghai Jiao Tong University, Shanghai, 200240, China}

%\affiliation{\shKeyLab}
\author{Zhou Huang}\affiliation{\shKeyLab}
\author{Guofang Shen}\affiliation{\BUAA}
\author{Qiuhong Wang}\affiliation{\FDU}
\author{Abdusalam Abdukerim}
\author{Zihao Bo}
\author{Wei Chen}\affiliation{\shKeyLab}
\author{Xun Chen}\affiliation{\shKeyLab}\affiliation{\SJTUSC}
\author{Yunhua Chen}\affiliation{\YaLongSD}
\author{Chen Cheng}\affiliation{\SYU}
\author{Yunshan Cheng}\affiliation{\SDUdep}\affiliation{\SDUlab}
\author{Xiangyi Cui}\affiliation{\TDLee}
\author{Yingjie Fan}\affiliation{\NKU}
\author{Deqing Fang}
\author{Changbo Fu}\affiliation{\FDU}
\author{Mengting Fu}\affiliation{\pku}
\author{Lisheng Geng}\affiliation{\BUAA}\affiliation{\BUAALab}\affiliation{\zzu}
\author{Karl Giboni}
\author{Linhui Gu}\affiliation{\shKeyLab}
\author{Xuyuan Guo}\affiliation{\YaLongSD}
\author{Chencheng Han}\affiliation{\shKeyLab}
\author{Ke Han}\affiliation{\shKeyLab}
\author{Changda He}\affiliation{\shKeyLab}
\author{Jinrong He}\affiliation{\YaLongSD}
\author{Di Huang}\affiliation{\shKeyLab}
\author{Yanlin Huang}\affiliation{\USST}

\author{Ruquan Hou}\affiliation{\SJTUSC}
\author{Xiangdong Ji}\affiliation{\UMD}
\author{Yonglin Ju}\affiliation{\MESJTU}
\author{Chenxiang Li}\affiliation{\shKeyLab}
\author{Mingchuan Li}\affiliation{\YaLongSD}
\author{Shu Li}\affiliation{\MESJTU}
\author{Shuaijie Li}\affiliation{\TDLee}
\author{Qing Lin}\affiliation{\USTClab}\affiliation{\USTCdep}
\author{Jianglai Liu}\email[Spokesperson: ]{jianglai.liu@sjtu.edu.cn}\affiliation{\shKeyLab}\affiliation{\TDLee}\affiliation{\SJTUSC}
\author{Xiaoying Lu}\affiliation{\SDUdep}\affiliation{\SDUlab}
\author{Lingyin Luo}\affiliation{\pku}
\author{Wenbo Ma}\affiliation{\shKeyLab}
\author{Yugang Ma}\affiliation{\FDU}
\author{Yajun Mao}\affiliation{\pku}
\author{Yue Meng}\affiliation{\shKeyLab}\affiliation{\SJTUSC}
\author{Xuyang Ning}\affiliation{\shKeyLab}
\author{Ningchun Qi}\affiliation{\YaLongSD}
\author{Zhicheng Qian}\affiliation{\shKeyLab}
\author{Xiangxiang Ren}\affiliation{\SDUdep}\affiliation{\SDUlab}
\author{Nasir Shaheed}\affiliation{\SDUdep}\affiliation{\SDUlab}
\author{Changsong Shang}\affiliation{\YaLongSD}

\author{Lin Si}\affiliation{\shKeyLab}
\author{Wenliang Sun}\affiliation{\YaLongSD}
\author{Andi Tan}\affiliation{\UMD}
\author{Yi Tao}\affiliation{\shKeyLab}\affiliation{\SJTUSC}
\author{Anqing Wang}\affiliation{\SDUdep}\affiliation{\SDUlab}
\author{Meng Wang}\affiliation{\SDUdep}\affiliation{\SDUlab}

\author{Shaobo Wang}\affiliation{\shKeyLab}\affiliation{\SPEIT}
\author{Siguang Wang}\affiliation{\pku}
\author{Wei Wang}\affiliation{\SYU}
\author{Xiuli Wang}\affiliation{\MESJTU}
\author{Zhou Wang}\affiliation{\shKeyLab}\affiliation{\SJTUSC}\affiliation{\TDLee}
\author{Mengmeng Wu}\affiliation{\SYU}
\author{Weihao Wu}
\author{Jingkai Xia}\affiliation{\shKeyLab}
\author{Mengjiao Xiao}\affiliation{\UMD}
\author{Xiang Xiao}\affiliation{\SYU}
\author{Pengwei Xie}\affiliation{\TDLee}
\author{Binbin Yan}\affiliation{\shKeyLab}
\author{Xiyu Yan}\affiliation{\USST}
\author{Jijun Yang}
\author{Yong Yang}\affiliation{\shKeyLab}
\author{Chunxu Yu}\affiliation{\NKU}
\author{Jumin Yuan}\affiliation{\SDUdep}\affiliation{\SDUlab}
\author{Ying Yuan}\affiliation{\shKeyLab}
\author{Dan Zhang}\affiliation{\UMD}
\author{Minzhen Zhang}\affiliation{\shKeyLab}
\author{Peng Zhang}\affiliation{\YaLongSD}
\author{Tao Zhang}
\author{Li Zhao}\affiliation{\shKeyLab}
\author{Qibin Zheng}\affiliation{\USST}
\author{Jifang Zhou}\affiliation{\YaLongSD}
\author{Ning Zhou}\email[Corresponding author: ]{nzhou@sjtu.edu.cn}\affiliation{\shKeyLab}
\author{Xiaopeng Zhou}\affiliation{\BUAA}
\author{Yong Zhou}\affiliation{\YaLongSD}

\collaboration{PandaX-4T Collaboration}
\noaffiliation

\date{\today}% It is always \today, today,

\begin{abstract}
Neutron-induced nuclear recoil background is critical to the dark matter searches in the PandaX-4T liquid xenon experiment. 
%This paper studies the feature of neutron background in liquid xenon and evaluates their contribution in the single scattering nuclear recoil events through Monte Carlo simulation and data-driven methods. 
This paper studies the feature of neutron background in liquid xenon and evaluates their contribution in the single scattering nuclear recoil events through three methods.
The first method is fully Monte Carlo simulation based. 
The last two are data-driven methods that also use the multiple scattering signals and high energy signals in the data, respectively.
In the PandaX-4T commissioning data with an exposure of 0.63 tonne-year, 
all these methods give a consistent result
that there are $1.15\pm0.57$ neutron-induced background in dark matter signal region
within an approximated nuclear recoil energy window between 5 and 100 keV.
\end{abstract}
\maketitle

%\linenumbers

\section{Introduction}

Dark matter is one of the top mysteries in modern physics~\cite{Plank_newest_result}.
Weakly interacting massive particle (WIMP) is one promising candidate, 
which is predicted by many new physics models beyond the Standard Model like super-symmetric theories. 
Searches on WIMPs have been performed with various experimental approaches for decades~\cite{jianglai_xun_xiangdong_review,zhaolireview}. 
The PandaX experiment, located in China Jinping Underground Laboratory (CJPL)~\cite{CJPL_intro,CJPL_intro_JNE,CJPL2_intro},
uses liquid xenon as the target to search for the scattering of WIMPs with xenon nuclei~\cite{finalcpc,Xia:2018qgsPandaXII,PandaX-II:2020udv,PandaX-II:2021nsg}.
Recently, the multi-ton scale stage of PandaX experiment, PandaX-4T, 
completed the commissioning run and released its first physics result~\cite{pandax4_first_paper}, 
which provides the world-leading constraints on the WIMP-nucleon spin-independent scattering cross section
for WIMP mass above 10~GeV/$c^2$.

The PandaX-4T detector is a cylindrical dual-phase xenon time projection chamber (TPC).
It is sequentially enveloped in an inner cryogenic pressure vessel, an outer pressure vessel, 
%and a 13~m ultra-pure water tank which are all made of stainless steel (SS).
both made of stainless steel (SS).
The outer pressure vessel is immersed in an ultrapure water shielding tank.
Two arrays of three-inch photomultipliers tubes (R11410 PMTs) are mounted on the top and at the bottom of the TPC, respectively.
Four electrodes (anode, gate, cathode and bottom screen) are placed in the TPC to provide the electric fields.
The anode and bottom screen electrodes are grounded 
and the gate and cathode electrodes are loaded with different negative high voltages.
The TPC is surrounded by polytetrafluoroethylene (PTFE) panels for higher photo collection efficiencies, 
which divide the detector into several regions together with four electrodes.
Out of the PTFE panels is the veto region for multi-scattering 
background rejection where one-inch PMTs are installed~\cite{pandax4t_sensitivity_paper}.
The sensitive volume
is confined within the PTFE panels, the gate and cathode electrodes, containing 3.7 tonne liquid xenon. 
The compartment between the cathode and the bottom screen electrodes is the below-cathode region.
An incident particle can generate a prompt scintillation signal
$S1$ and a delayed electroluminescence signal $S2$ in the TPC,
%q$S1$ and q$S2$ denote the charge of these signals.
which are collected to reconstruct the event position and deposited energy~\cite{pandax4_first_paper}.

A robust evaluation of the neutron background contribution is crucial for the dark matter searches.
A WIMP particle is expected to scatter off the xenon nucleus elastically, 
which releases a single scattering nuclear recoil (SSNR) signal. 
However, neutrons, generated from the detector materials via ($\alpha$,n) or spontaneous fission (SF) reaction~\cite{alpha_n_F19,alpha_n_production},
can also produce NR signals, mimicking WIMP particles. 
Based on the measured radioactivities of all the detector components,
a Monte Carlo (MC) simulation can predict the neutron yields and their energy deposits in liquid xenon. However, this method relies on the accuracy of radioacitivity measurement and single scattering reconstruction algorithm in the TPC.
Different from WIMPs, neutrons have \textcolor{black}{some} probabilities of scattering multiple times
and finally getting captured by the xenon nuclei.
If capturing the neutron, 
the xenon nuclei would produce high energy gamma (HEG) rays through de-excitation.
With these features, multiple scattering nuclear recoil (MSNR) events and HEG events in the data can be used to estimate the amount of neutron-induced SSNR contribution.

The aim of this paper is to describe the process of neutron-induced SSNR background evaluation in PandaX-4T experiment.
The rest of this paper is organized as follows.
In Sec.~\ref{chapter:response}, the NR calibration sources and detector response are introduced.
In Sec.~\ref{chapter:MC_simulation}, 
the MC simulation processing and SSNR background evaluation from the simulation are presented.
The details of the two data-driven methods are introduced in Sec.~\ref{chapter:data_driven_method}, 
followed by a summary in Sec.~\ref{chapter:N_estimation}.

\section{Detector response to neutrons}
\label{chapter:response}

The detector response to the neutron-induced nuclear recoil (NR) events is modeled and validated through neutron calibration data.
During the commissioning run of PandaX-4T, 
the $\rm ^{241}Am-Be$ (AmBe) and deuteron-deuteron (DD) neutron sources are deployed outside the TPC to generate NR events.

The AmBe source is a widely used NR calibration source in dark matter direct detection experiments.
The $\alpha$ particles are emitted from $^{241}$Am and get captured by $^{9}$Be, generating neutrons consequently:
\begin{eqnarray}
^{9}\rm{Be}+\alpha \to \rm{n}+^{12}\rm{C}^*\,.\nonumber
\end{eqnarray}
This is the so-called ($\alpha$, n) reaction~\cite{sources4a_code,alpha_n_production}.
The final product $^{12}$C primarily lies in the ground state or the first excited state. 
The latter case can emit a 4.4~MeV gamma ray.
PandaX-4T calibration uses the same AmBe source as PandaX-II~\cite{p2neutron_paper,pandax2calibration}. 
The source is movable inside the calibration polyethylene tubes 
between the outer vessel and inner cryogenic vessel~\cite{pandax4t_cryo_design,pandax_SS_vessel}, 
and was placed at some certain positions during the calibration data taking.
\textcolor{black}{
The commissioning data was divided into 5 data sets according to different conditions~\cite{pandax4_first_paper}.
}
Throughout the commissioning, two periods of AmBe calibration were performed, one between data set~3 and data set~4, the other at the end of data set~5, accumulating a calibration time of 174.2 hours in total~\cite{pandax4_first_paper}.

Monoenergetic neutrons can be generated from DD nuclear collision, 
\begin{eqnarray}
    \rm{D} + \rm{D} \to ~^{3}\rm{He} + \rm{n}\,.\nonumber
\end{eqnarray}
The kinetic energy of the product neutron depends on the deuteron energy and the emission angle~\cite{LUX_DDcalibration,DDgenerator_paper,csikai1987}.
In PandaX-4T experiment, the DD generator is placed near the outer surface of ultrapure water shielding tank, 
where a stainless steel pipeline is welded.
The generated neutrons are guided by the pipeline and point to the center of the PandaX-4T detector.
Two sets of DD NR calibration data were collected after data set~5, 
with 2.2~MeV and 2.45~MeV neutron energy respectively.
These two conditions correspond to $\pi$ and $\pi$/2 neutron emission angles.
\textcolor{black}{
In total, the DD calibration live time is 86.0~hours.
}

From NR calibration data, the SSNR events with only one good pair of $S1$ and $S2$ are selected to develop 
the fiducial volume (FV) cut,
the veto cut, several quality cuts and the corresponding efficiencies as described in Ref.~\cite{pandax4_first_paper}.
These cuts and efficiencies are also necessary for SSNR background evaluation.
The electron-equivalent energy $E$ (unit keV$_{\rm ee}$) is reconstructed 
from q$S1$(the charge of $S1$ signal in the unit of photoelectron, PE)
and q$S2_{\rm b}$(the scintillation in $S2$ collected by the bottom PMTs in the unit of PE).
The same detector parameters in Ref.~\cite{pandax4_first_paper}, 
photon detection efficiency(PDE), 
electron extraction efficiency(EEE) and single electron gain(SEG$_{\rm b}$),
are adopted.
%Also the cut efficiency is obtained and applied to the NR calibration simulation.
In total, 2721 AmBe SSNR events and 2606 DD SSNR events are selected in the energy range of [0, 20] keV$_{\rm ee}$.
Based on these SSNR events and simulation, the signal response model is constructed 
and agrees with data within uncertainties~\cite{pandax4_first_paper}. 

In addition, the MSNR events and neutron-capture events in the calibration data provide validation to the simulation,
which will be described \textcolor{black}{later}.
%in Sec.~\ref{chapter:MC_simulation} and Sec.~\ref{chapter:data_driven_method}.

\section{Monte Carlo Method}
\label{chapter:MC_simulation}
The Monte Carlo (MC) method can simulate the decay of radioactive isotopes in the detector materials 
and the consequent energy deposit in liquid xenon from the decay products 
including neutrons, gammas, beta and alpha particles, etc~\cite{pandax4t_sensitivity_paper,qzcmaterial}. 
The radioactivities of the main materials,
including stainless steel(SS), PTFE, PMTs and the readout bases,
are measured with the high-purity germanium detector and shown in Tab.~\ref{table:radioactivity}.
There are some differences as compared with Ref.~\cite{pandax4t_sensitivity_paper},
which results from the treatment that the radioactivity has been updated according
to real materials used in PandaX-4T experiment.
Especially, the PMT shell was not considered in previous work~\cite{pandax4t_sensitivity_paper,p2neutron_paper}. 
The neutron yields and the neutron energy spectra of ($\alpha$,n) and SF reactions
are calculated by SOURCES-4A code~\cite{sources4a_code}. 
The neutron yields are summarized in Tab.~\ref{table:neutron_yield}. 
With the above inputs, a full detector MC simulation, utilizing the GEANT4 package~\cite{geant4_collaboration}, 
gives the neutron background estimation.
The disequilibrium in the $^{238}$U and $^{232}$Th chains is considered in the simulation.
The $^{238}$U$_{\rm e}$ represents the early chain of $^{238}$U decay, 
which ends at $^{230}$Th. 
The $^{238}$U$_{\rm l}$ represents the late chain of $^{238}$U decay, 
which starts from $^{226}$Ra and ends at $^{206}$Pb. 
The $^{232}$Th$_{\rm l}$ represents the late chain of $^{232}$Th decay, 
which starts from $^{228}$Th and ends at $^{208}$Pb. 
The early chain of $^{232}$Th is not considered as the neutron yield from this part is negligible~\cite{p2neutron_paper}.

%radioactivity table
\begin{table}[htbp]
\renewcommand\arraystretch{1.2}
\footnotesize
\caption{Radioactivity of the materials used in this analysis. 
The different components of cryostat (two vessels) are sampled and measured separately, 
as shown in the first four rows.}
\doublerulesep 0.1pt \tabcolsep 10pt
\centering
\begin{tabular}{cc|cccc}
\hline\hline
\multirow{2}{*}{Component} & \multirow{2}{*}{Quantity} & \multicolumn{4}{c}{Radioactivity (mBq/kg or mBq/piece)} \\ \cline{3-6} 
                           &                           & $^{235}$U   & $^{232}$Th$_{\rm l}$   & $^{238}$U$_{\rm e}$    & $^{238}$U$_{\rm l}$       \\ \hline

Inner Vessel barrel and dome (SS)   & 443.5 kg  & $0.32 \pm 2.67$ & $2.54 \pm 1.82$ & $30.23 \pm 41.16$ & $3.21 \pm 2.04$   \\
Outer Vessel barrel (SS)            & 961.4 kg  & $5.26 \pm 2.71$ & $3.17 \pm 1.83$ & $40.87 \pm 22.77$ & $1.97 \pm 1.44$   \\
Outer Vessel dome (SS)              & 396.6 kg  & $2.78 \pm 2.42$ & $4.89 \pm 1.72$ & $40.84 \pm 24.03$ & $2.84 \pm 1.33$   \\
Flange (SS)                         & 1254.5 kg & $2.81 \pm 1.90$ & $4.42 \pm 1.82$ & $0.00  \pm 15.81$ & $1.81 \pm 1.33$   \\
R11410 PMT                      & 368 pieces    & $4.60 \pm 8.53$ & $2.46 \pm 0.96$ & $26.29 \pm 16.90$ & $3.23 \pm 1.18$   \\
PMT base                        & 368 pieces    & $0.46 \pm 1.22$ & $0.28 \pm 0.18$ & $6.97  \pm 1.94 $ & $0.84 \pm 0.22$   \\
PTFE        & 200.0 kg                          & $(0.10 \pm 0.10)\times 10^{-3}$ & $0.04 \pm 0.04$ & $0.02 \pm 0.02$ & $0.02 \pm 0.02$   \\
\hline\hline
\end{tabular}
\label{table:radioactivity}
\end{table}

%neutron yield table
\begin{table}[htbp]
\renewcommand\arraystretch{1.2}
\footnotesize
\caption{Neutron yield of radioactive decay chain for different materials in the unit of neutron/decay.
    The SF is dominated by the $^{238}$U isotope.}
\doublerulesep 0.1pt \tabcolsep 10pt
\centering
\begin{tabular}{c|cccc|c}
\hline\hline
\multirow{2}{*}{Component} & \multicolumn{4}{c|}{($\alpha$,n)} & SF    \\   \cline{2-6}
                           & $^{235}$U  & $^{232}$Th$_{\rm l}$   & $^{238}$U$_{\rm e}$  & $^{238}$U$_{\rm l}$   & $^{238}$U$_{\rm e}$  \\  \hline
SS (cryostat)   & 3.4$\times 10^{-7}$ & 1.5$\times 10^{-6}$  &8.3$\times 10^{-10}$   & 4.1$\times 10^{-7}$ & 1.1$\times 10^{-6}$  \\
PTFE      & 8.9$\times 10^{-5}$ & 8.7$\times 10^{-5}$  & 9.4$\times 10^{-6}$   & 5.2$\times 10^{-5}$   & 1.1$\times 10^{-6}$      \\
Kovar (PMT shell)   &1.6$\times 10^{-7}$ & 9.0$\times 10^{-7}$ & 7.6$\times 10^{-11}$ & 2.2$\times 10^{-7}$ & 1.1$\times 10^{-6}$  \\
SiO$_2$ (PMT window)&1.5$\times 10^{-6}$ &1.6$\times 10^{-6}$ &9.5$\times 10^{-8}$ &9.6$\times 10^{-7}$ &1.1$\times 10^{-6}$ \\
Cirlex (PMT base)  & 2.2$\times 10^{-6}$ &2.3$\times 10^{-6}$ & 3.5$\times 10^{-7}$ & 1.4$\times 10^{-6}$ & 1.1$\times 10^{-6}$ \\
SS (PMT electrode)     & 3.4$\times 10^{-7}$ & 1.5$\times 10^{-6}$ &8.3$\times 10^{-10}$ & 4.1$\times 10^{-7}$ & 1.1$\times 10^{-6}$     \\
Al$_2$O$_3$ (PMT ceramic)  & 9.5$\times 10^{-6}$ & 1.1$\times 10^{-5}$ & 2.7$\times 10^{-7}$ &6.2$\times 10^{-6}$   &1.1$\times 10^{-6}$ \\
\hline\hline
\end{tabular}
\label{table:neutron_yield}
\end{table}

\begin{figure}[htbp]
\centering
\includegraphics[width=0.5\columnwidth]{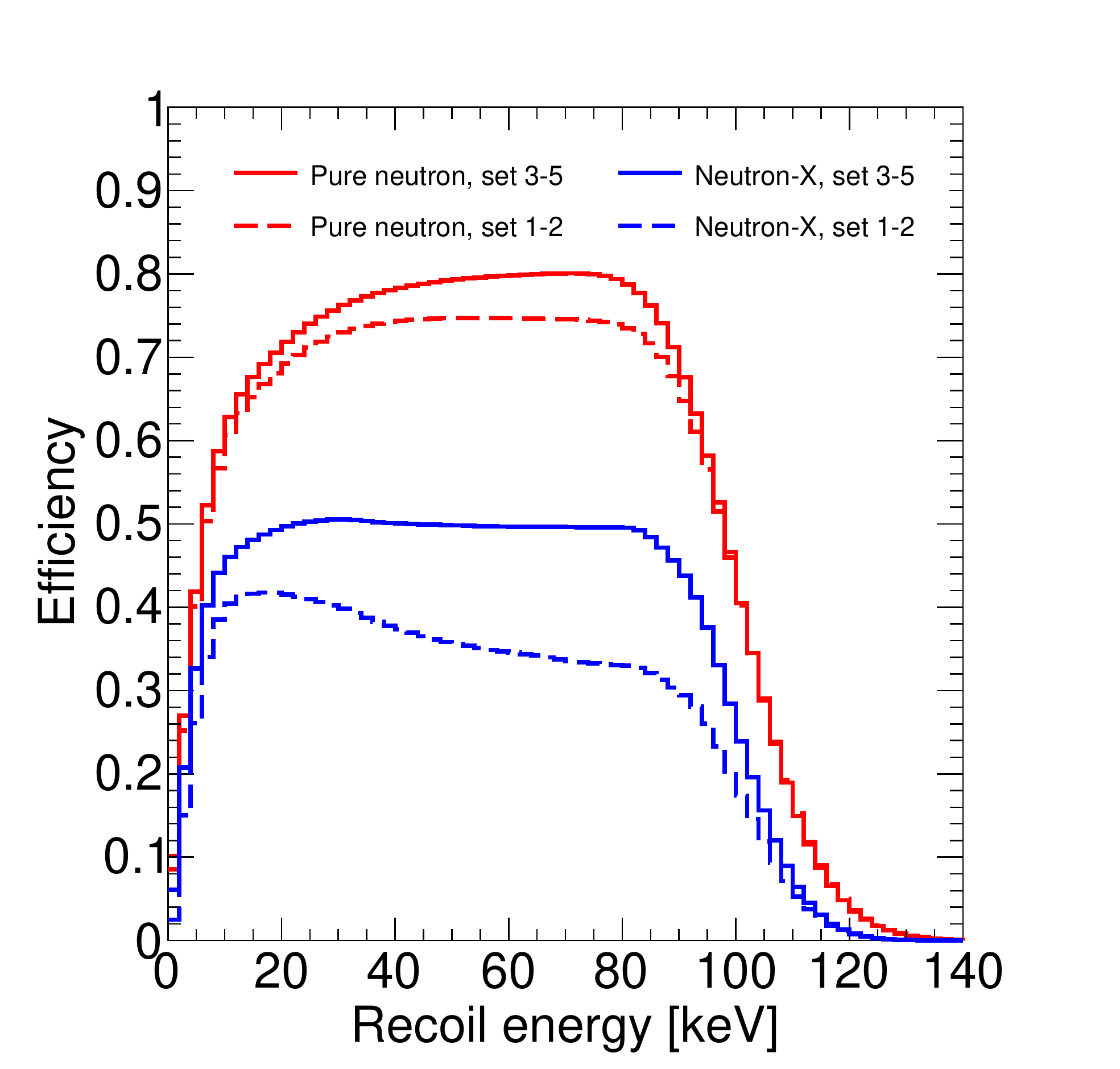}
\caption{Selection efficiency as a function a nuclear recoil energy from signal model. 
Dashed and solid curves represent the efficiencies in the different data sets. 
Pure neutron and neutron-X are the two types of neutron-induced SSNR background.
\textcolor{black}{
The region of interest cuts have selection boundaries, i.e. qS1 $<$ 400~PE and qS2 $<$ 20000~PE. 
It leads to the efficiency curves decrease toward higher energy.
The cuts are kept consistent over different data sets
while the signal yields are different.
The NR acceptance cut make the efficiency of data sets 1-2 lower than that of other data sets.
Also, it leads to the smaller efficiency for neutron-X.}
}
\label{fig:efficiency}
\end{figure}

The detector simulation records the number of scattering,
the deposited energy and the position information of each deposition. Several improvements have been adopted in this work, including a more detailed geometry, a more realistic veto energy cut, and the fiducial volume (FV) cut of PandaX-4T commissioning run.  The veto cut threshold in simulation is adjusted to 705 keV$_{\rm ee}$ by comparing the data and MC.
Due to the lower detection efficiency in the veto region, 
this threshold is higher than that in PandaX-II~\cite{pandax4t_sensitivity_paper}.
In PandaX-4T experiment,
it provides approximately 20\% reduction in the SSNR event. The estimated neutron-induced SSNR event contribution is formalized as
\begin{eqnarray}
	 N_{{\rm ssnr}} &=& \sum_{i}\sum_{j} \left( A_{ij} \times Y_{ij} \times M_i \times P_{{\rm ssnr},ij} \right) \times \epsilon  \times T,
\end{eqnarray}
where $M_i$ is the mass of the material component $i$, $A_{ij}$ is the measured radioactivity of the isotope $j$ in the material $i$,
$Y_{ij}$ is the neutron yield.
$P_{{\rm ssnr},ij}$ is the probability that a neutron leads to the final SSNR background, 
which is based on detector MC simulation.
The detection efficiency, $\epsilon$, is different among date sets and types, as shown in Fig.~\ref{fig:efficiency}.
$T$ is the duration of dark matter searching. 

There are some fraction of neutrons depositing part of their energy in the below-cathode region and part in the active volume. 
Due to the inverse field below cathode, 
the ionized electrons are unable to drift upward to the gas xenon region and produce $S2$ signals. 
However, the prompt $S1$ signals are collected. 
These events can also contribute to the SSNR background, 
and are called ``neutron-X'' events. 
PandaX-4T detector is not capable of distinguishing them with ``pure neutron'' events, 
that neutrons only deposit their energy in the active volume.
These two kinds of SSNR events have different distributions in the $\log_{10} ({\rm q}S2_{\rm b}/{\rm q}S1)$ versus q$S1$ parameter space, 
which is critical for the background fitting~\cite{pandax4_first_paper}. 
In the simulation, the neutron-X events can be identified, and their contribution can be calculated separately.

%Table~\ref{table:MS_multiplicity} shows the number of feature events selected in calibration data as compared with MC simulation. 
%In the simulation, 
%the energy deposits are grouped in a consistent way to the data clustering algorithm, 
%then neutron events are classified according to the number of grouped deposits (SSNR and MSNR). 
%In addition, the associated HEG events from neutron capture are simulated.
%The MC predicted MSNR-to-SSNR ratio is consistent with the data within 33\%, 
%which is mainly due to the inaccurate handling on the deposition clustering in MC and taken as one systematic uncertainty.

The simulation is validated with AmBe and DD calibration data. 
In the simulation, the energy deposits are grouped in the same way as the data clustering algorithm, 
then neutron events are classified according to the number of grouped deposits (SSNR and MSNR). 
In addition, the associated high energy gammas (HEG) from neutron capture are simulated. 
The selection of these MSNR and HEG events are described in Sec.~\ref{chapter:data_driven_method}. 
Table~\ref{table:MS_multiplicity} gives the comparison of MSNR-to-SSNR and HEG-to-SSNR ratios between data and MC simulation. 
The largest difference, 33\%, is adopted as the simulation uncertainty.

%feature event numbers in DD and AmBe
\begin{table*}[htbp]
\renewcommand\arraystretch{1.2}
\footnotesize
\caption{Event number comparison between NR calibration data and MC simulation.
The SSNR event, MSNR events, HEG events and their relative ratio $R$ are listed.
Forty million DD neutron events and twenty-eight million AmBe neutron events are generated in simulation.
\textcolor{black}{
The event numbers in the MC are normalized to the SSNR event numbers in data.
}
Noted that the SSNR events in this table are within the energy range of [0-20]~keV$_{\rm ee}$.
Also, the event selection volumes for SSNR, MSNR and HEG are not same (introduced later),
but are all consistent in data and MC simulation.
}
\doublerulesep 0.1pt \tabcolsep 10pt
\centering
\begin{tabular}{c|cc|cc}
\hline\hline
Event count & DD data   & DD MC     & AmBe data         & AmBe MC   \\ \hline
SSNR        & 2606      & 2606      & 2721              & 2721     \\
MSNR (N=2)  & 3340      & 3444      & 3082              & 3489     \\
MSNR (N=3)  & 1953      & 2335      & 1541              & 2313     \\
MSNR (N=4)  & 1138      & 1439      & 751               & 1527     \\
MSNR (N=5)  & 519       & 858       & 397               & 1019     \\
MSNR (N$\le$5) & 6950   & 8077      & 5771              & 8350    \\ 
HEG         & 71457     & 75036     & 98450             & 93464   \\ \hline
R=MSNR/SSNR   & 2.7     & 3.1       & 2.12              & 3.1       \\
R=HEG/SSNR    & 27.4    & 28.8      & 36.2              & 34.3      \\ 
\hline\hline
\end{tabular}
\label{table:MS_multiplicity}
\end{table*}

%MC neutron spectra
\begin{figure}[htbp]
\centering
\includegraphics[width=0.45\columnwidth]{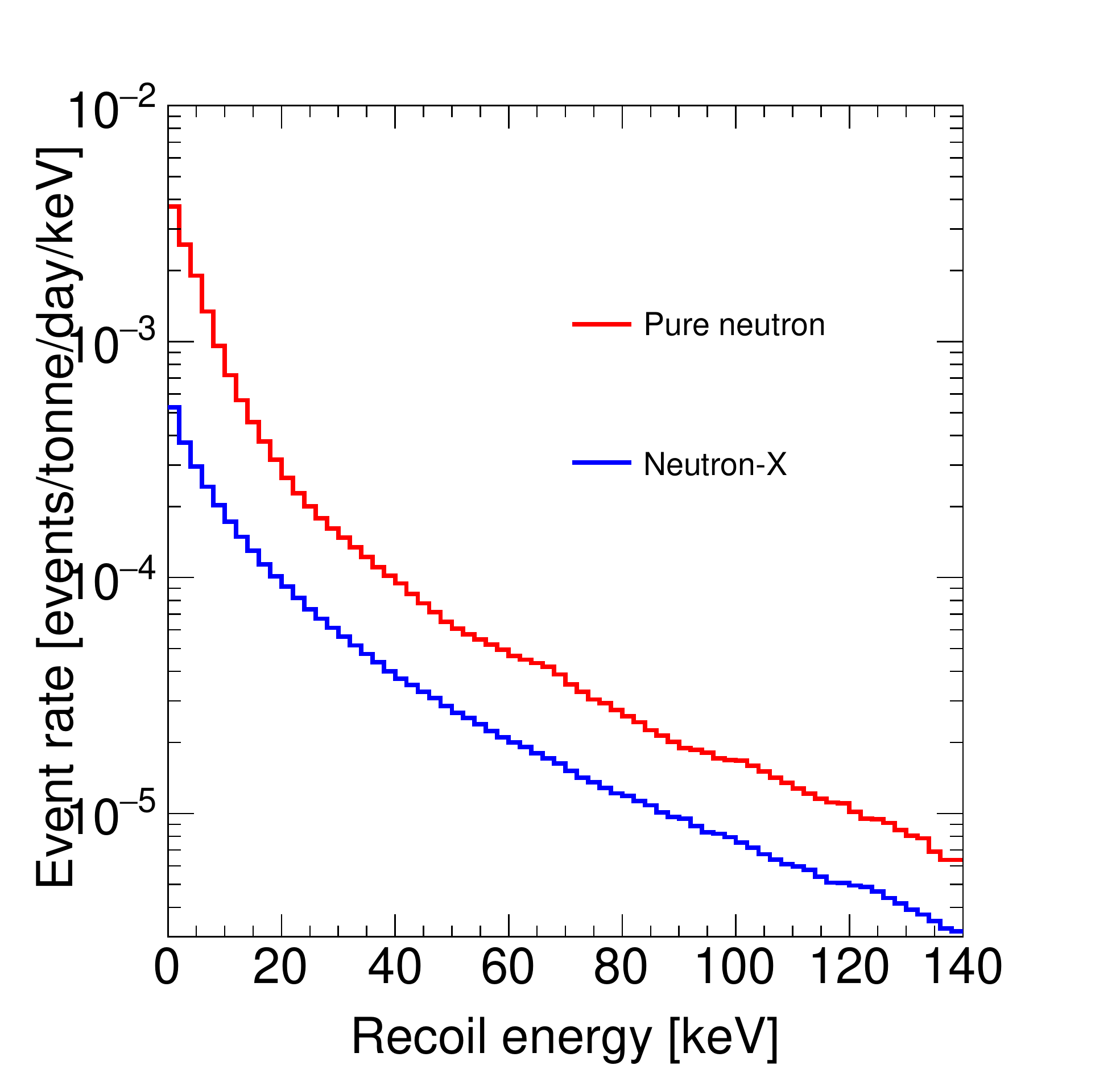}
\includegraphics[width=0.45\columnwidth]{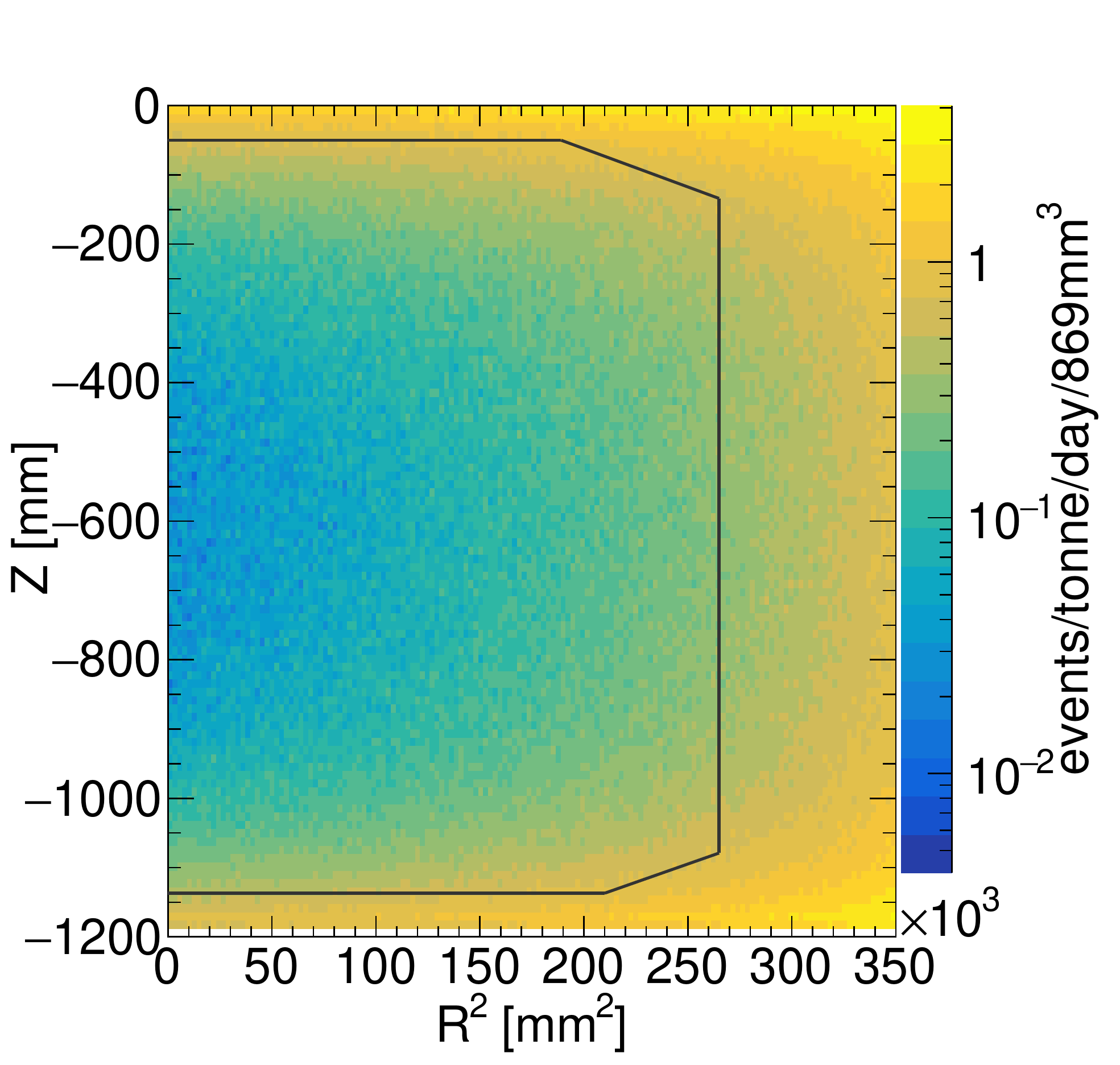}
\caption{
\textcolor{black}{
Left: The energy spectra for neutron background events inside FV.
        Right: The total spatial distribution for the neutron background events from MC simulation.
        The solid black lines represent the fiducial volume boundaries, resulting the FV mass of 2.67 tonne.}
        }
\label{fig:MC_Espectrum}
\end{figure}

%money table, event rate
\begin{table*}[htbp]
\renewcommand\arraystretch{1.2}
\centering
\footnotesize
\caption{Predicted event rate for different materials from the MC simulation.
Also the ratios between MSNR/HEG and pure neutron/neutron-X are listed in the last two rows.}
\doublerulesep 0.1pt %\tabcolsep 10pt
\centering
\begin{tabular}{c|cccc|cccc}
\hline\hline
Data sets               & \multicolumn{4}{c|}{Set 1-2}               & \multicolumn{4}{c}{Set 3-5}                \\ \hline
Duration         & \multicolumn{4}{c|}{14.0 days}                  & \multicolumn{4}{c}{72.0 days}                   \\ \hline
Rate (counts/day) & Pure neutron & Neutron-X & MSNR   & HEG    & Pure~neutron & Neutron-X & MSNR   & HEG    \\ \hline
Inner Vessel           & 2.1$\times 10^{-3}$       & 2.0$\times 10^{-4}$    & 1.0$\times 10^{-2}$ & 6.5$\times 10^{-2}$ & 2.2$\times 10^{-3}$       & 2.6$\times 10^{-4}$    & 1.0$\times 10^{-2}$ & 6.5$\times 10^{-2}$ \\
Outer Vessel           & 4.2$\times 10^{-3}$       & 4.6$\times 10^{-4}$    & 2.0$\times 10^{-2}$ & 1.4$\times 10^{-1}$ & 4.4$\times 10^{-3}$       & 5.9$\times 10^{-4}$    & 2.0$\times 10^{-2}$ & 1.4$\times 10^{-1}$ \\
PTFE                    & 1.4$\times 10^{-4}$       & 1.8$\times 10^{-5}$    & 1.0$\times 10^{-3}$ & 5.9$\times 10^{-3}$ & 1.5$\times 10^{-4}$       & 2.2$\times 10^{-5}$    & 1.1$\times 10^{-3}$ & 5.9$\times 10^{-3}$ \\
R11410 PMT   & 1.2$\times 10^{-3}$       & 6.2$\times 10^{-4}$    & 7.3$\times 10^{-3}$ & 3.1$\times 10^{-2}$ & 1.3$\times 10^{-3}$       & 7.8$\times 10^{-4}$    & 7.6$\times 10^{-3}$ & 3.1$\times 10^{-2}$ \\
PMT base    & 6.1$\times 10^{-4}$ & 3.5$\times 10^{-4}$ & 3.7$\times 10^{-3}$ & 2.3$\times 10^{-2}$ & 6.6$\times 10^{-4}$ & 4.4$\times 10^{-4}$ & 3.9$\times 10^{-3}$ & 2.3$\times 10^{-2}$ \\ \hline
Total                   & 8.2$\times 10^{-3}$       & 1.6$\times 10^{-3}$    & 4.1$\times 10^{-2}$ & 2.6$\times 10^{-1}$ & 8.7$\times 10^{-3}$       & 2.1$\times 10^{-3}$    & 4.3$\times 10^{-2}$ & 2.6$\times 10^{-1}$ \\ \hline
$R_{\rm MC}$~pure~neutron   & -          & -       & 5.0    & 32.1   & -          & -       & 5.0    & 30.3   \\
$R_{\rm MC}$~neutron-X      & -          & -       & 25.0   & 160.2  & -          & -       & 20.8   & 126.1  \\
\hline\hline
\end{tabular}
\label{table:N_events_MC}
\end{table*}

The energy spectra for pure neutron and neutron-X events,
and their total spatial distribution are shown in Fig.\ref{fig:MC_Espectrum}.
For the MC method, the uncertainty \textcolor{black}{most} comes from the radioactivity measurement and simulation.
%The $+1~\sigma$ upper limit of radioactivity gives around 37\% increment in the SSNR background events.
Based on Tab.~\ref{table:radioactivity}, the radioactivity measurement uncertainty is calculated, 37\%.
\textcolor{black}{
The uncertainty from the SOURCES4A calculation is around 17\%~\cite{sources4a_code}.
In total, the uncertainty of MC method is 52\%.
}
The predicted neutron-induced SSNR background events are 0.12~$\pm$~0.06 (pure neutron in data sets 1-2),
\textcolor{black}{0.62~$\pm$~0.32} (pure neutron in data sets 3-5), 0.02~$\pm$~0.01 (neutron-X in data sets 1-2),
\textcolor{black}{0.15~$\pm$~0.08} (neutron-X in data sets 3-5) and summarized in Tab.~\ref{table:N_events_final}.
\textcolor{black}{The detailed values from different component are listed in Tab.~\ref{table:N_events_MC}.}

\section{Data driven method}
\label{chapter:data_driven_method}

To better estimate the neutron background in the data, 
various data driven methods have been developed~\cite{xenon1t_analysis_paper2,p2neutron_paper}.
Besides the SSNR events, the MSNR events and HEG events
~\cite{neutron_bkg_estimation_2004,xenon1t_analysis_paper2,p2neutron_paper} are also the features of neutrons, which are distinguishable signals in the data.
Therefore, neutron background can be evaluated by 
\begin{eqnarray}\label{eqn:data_driven_formula}
    N_{\rm{ssnr}} &=& \frac{N_{\rm{feature}}}{ R_{\rm{MC}}}
\end{eqnarray}
where $N_{\rm{feature}}$ refers to the number of MSNR or HEG events, the $R_{\rm{MC}}$ is the ratio between the featured events and SSNR events obtained from simulation \textcolor{black}{(last two rows in Tab.~\ref{table:N_events_MC})}.

\subsection{Multi-Scatter}
\label{chapter:MSNR_selection}

The neutrons are likely to scatter multiple times in the PandaX-4T detector, and result in MSNR events.
The kinetic energy of fast neutrons is several MeV. 
The neutron velocity is $\sim 0.1~c$ and the mean free path is several centimeters~\cite{neutron_bkg_estimation_2004,neutron_mean_path}.
The average time separation of adjacent scattering is several nanoseconds. 
The prompt light travels meters in the PandaX-4T detector before reaching the PMTs, 
and the width of $S1$ signal can reach $\sim 100$ nanoseconds. 
Therefore, the $S1$ signals in MSNR events are reconstructed as a single $S1$. 
However, if multiple scatters occur at different vertical positions, 
which is true in most cases, these $S2$ signals do not overlap with each other and can be identified.
The horizontal position of each scatter is reconstructed through the light pattern of the top PMT array~\cite{p2_pos_rec_paper}.
The vertical positions are determined by the time separation between the $S2$s and the combined $S1$.

After 3-D uniformity correction applied on the $S1$ and $S2$s,
the combined electron-equivalent energy of the MSNR
event, $E_{\rm{MSNR}}$, is reconstructed,
which follows the SSNR energy reconstruction formula. 
\textcolor{black}{
The correction for the S1 signal is solely based on the position of the largest S2
and the resulting uncertainty for $\sum$q$S1$ is less than 10\%.
This uncertainty has been incorporated in the comparison between calibration data and MC.
}
The energy region of interest for MSNR is set as 1~keV$_{\rm{ee}}  < E_{\rm{MSNR}} <$ 
25~keV$_{\rm{ee}}$
and the number of scatters should be larger than one and less than six,
for the reason that they cover more than 80\% of the whole multi-scattering nuclear recoil events in data and MC simulation.
To collect more multiple scatter NR event, 
a larger fiducial volume (LFV) is defined as following, resulting in a target xenon mass of 3.04 tonne.
The position radius square of the scattering with the maximum $S2$, is confined within $3 \times 10^{5}$~mm$^2$.
The vertical position confinement of this maximum $S2$ follows the SSNR's, i.e., 
52~mm below the gate and 58~mm above the cathode.
There is no restriction on the positions of other scatterings.

\begin{figure}[htbp]
\includegraphics[width=0.45\columnwidth]{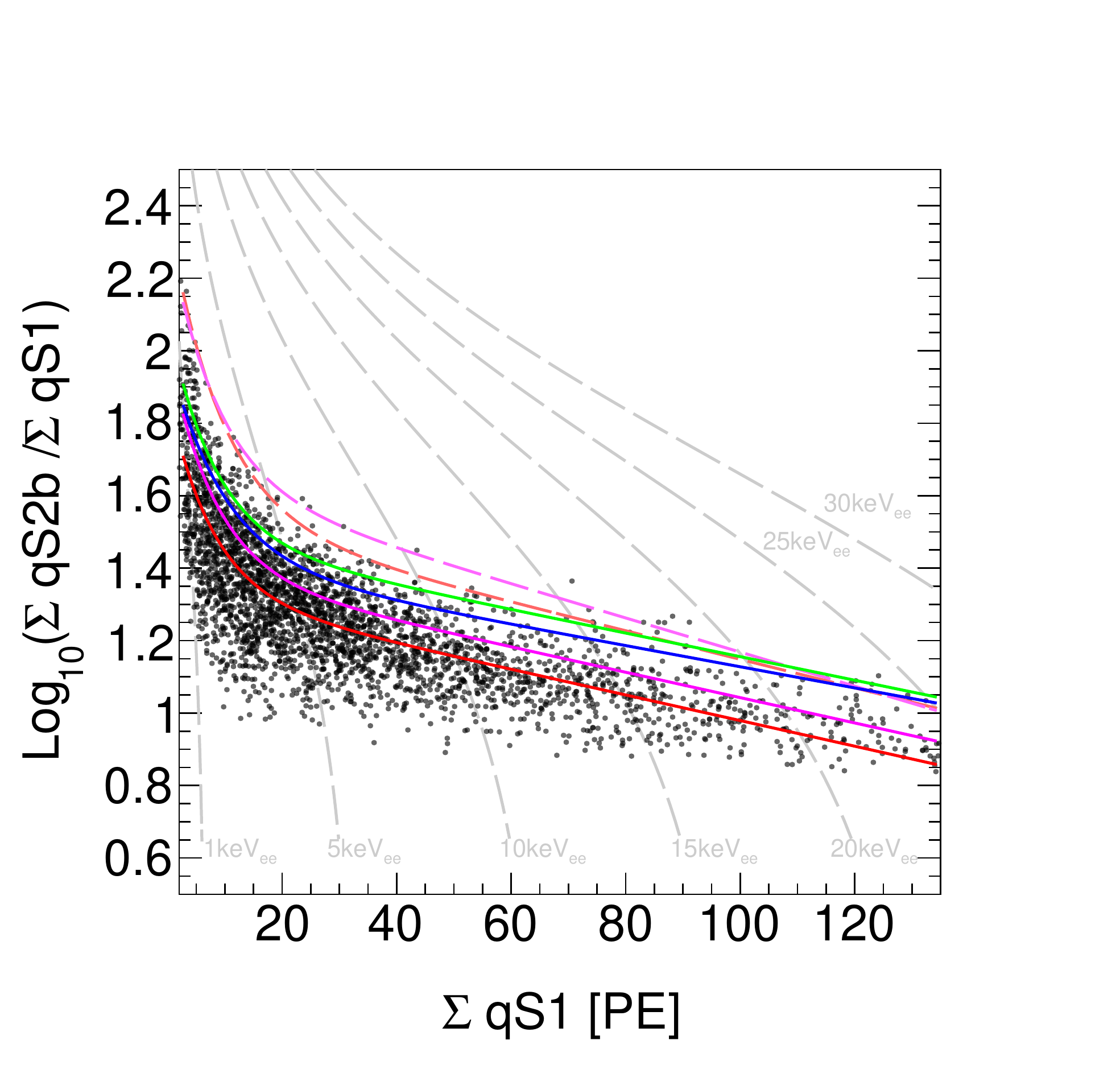}
\includegraphics[width=0.45\columnwidth]{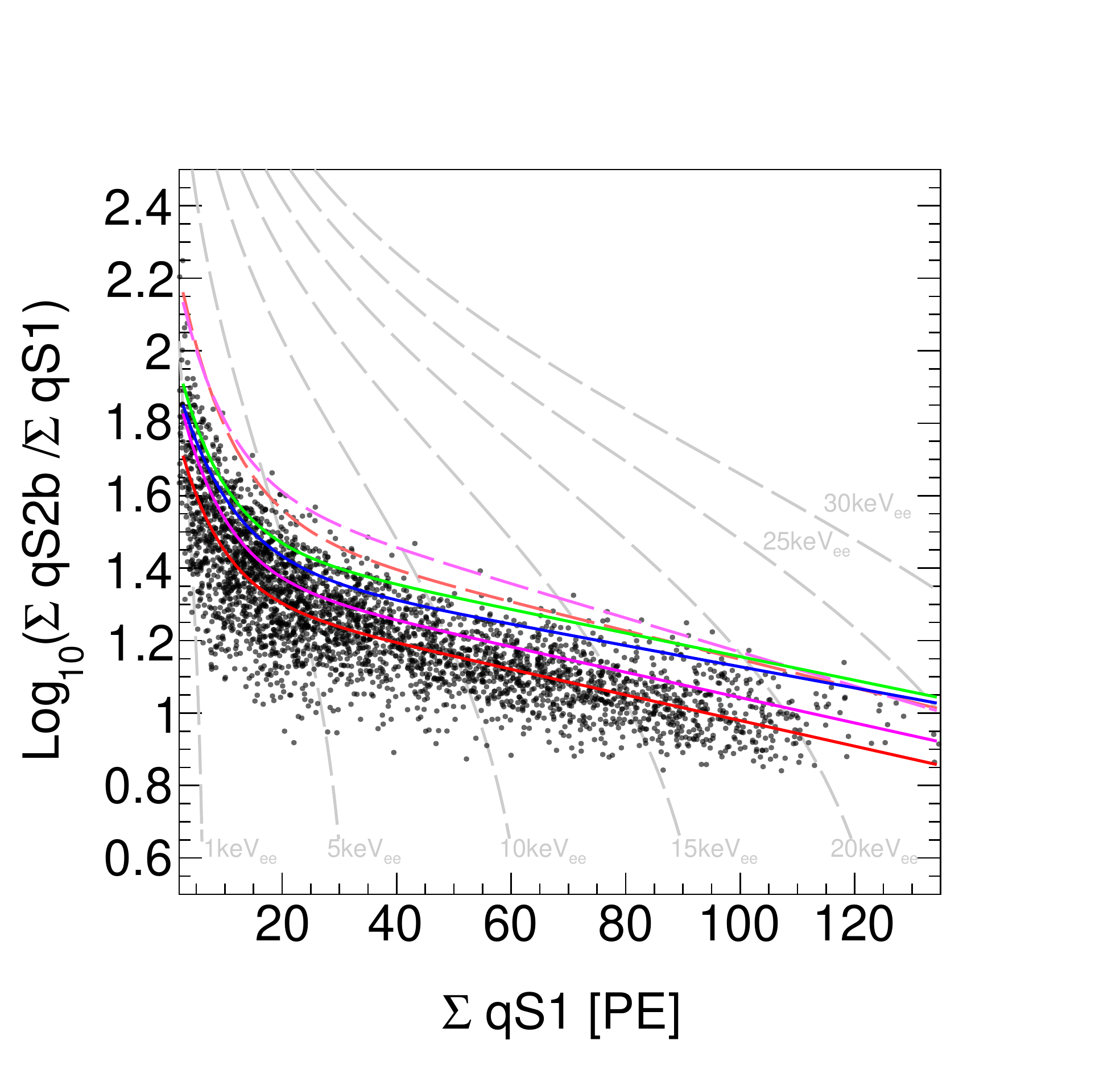}
\caption{
\textcolor{black}{
The distribution on $\log_{10}(\sum{{\rm q}S2_{\rm b}}/\sum{{\rm q}S1})$ versus $\sum{{\rm q}S1}$
of double scattering NR events in AmBe data (Left) and DD data (Right).
}
    The red, magenta, blue and green solid lines represent the DD and AmBe combined MSNR median curve for 2, 3, 4 and 5 scattering NR events, respectively.
    The red and magenta dashed lines show the 99\% upper quantile boundaries for 2 and 3 scattering NR events.
    The electron equivalent recoil energy in keV$_{\rm ee}$ is indicated with the grey dot dashed lines.
    A clear shift of the median curves are shown,
    which is typical for the multiple scattering events.}
\label{fig:MS_log10S2S1_versus_S1}
\end{figure}

Figure~\ref{fig:MS_log10S2S1_versus_S1} shows the double scattering NR event distribution 
of $\log_{10}(\sum{{\rm q}S2_{\rm b}}/\sum{{\rm q}S1})$ versus $\sum{{\rm q}S1}$ from AmBe and DD calibration data.
The 99\% upper quantiles of this distribution can be derived 
and are also plotted.
These so-called 99\% acceptance cuts are applied 
to the MSNR candidates selection to suppress the multiple scattering electron recoil (ER) events.
For number of scattering larger than 3, the statistic is not enough.
However, the ER MC simulation shows that 
the multiple-scatter ER contamination is negligible in this region.
Thus the 99\% acceptance cuts are not necessary.

%MSNR logS2S1 distribution

\begin{figure}[htbp]
  \centering
  \includegraphics[width=0.45\columnwidth]{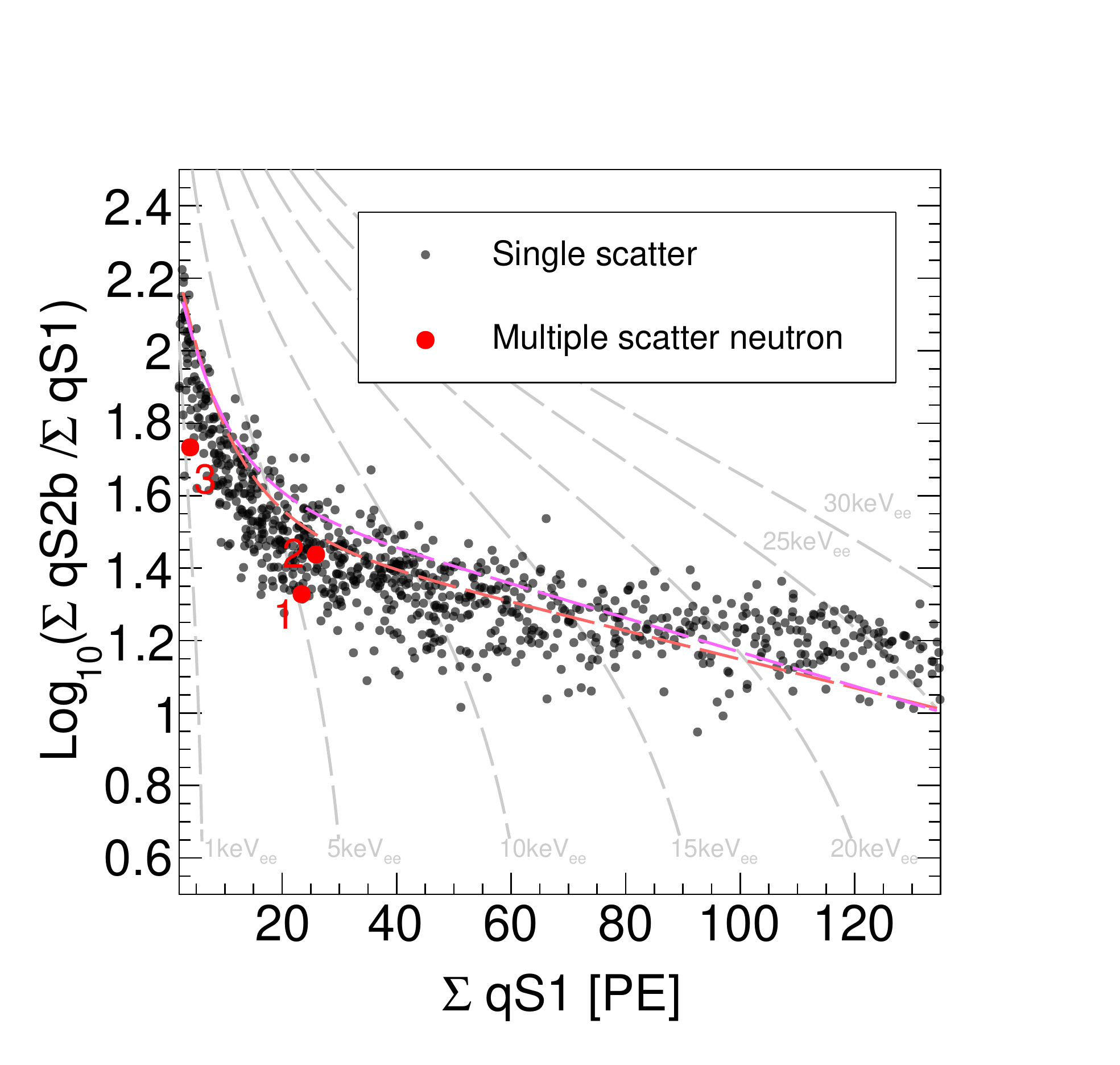}
  \includegraphics[width=0.45\columnwidth]{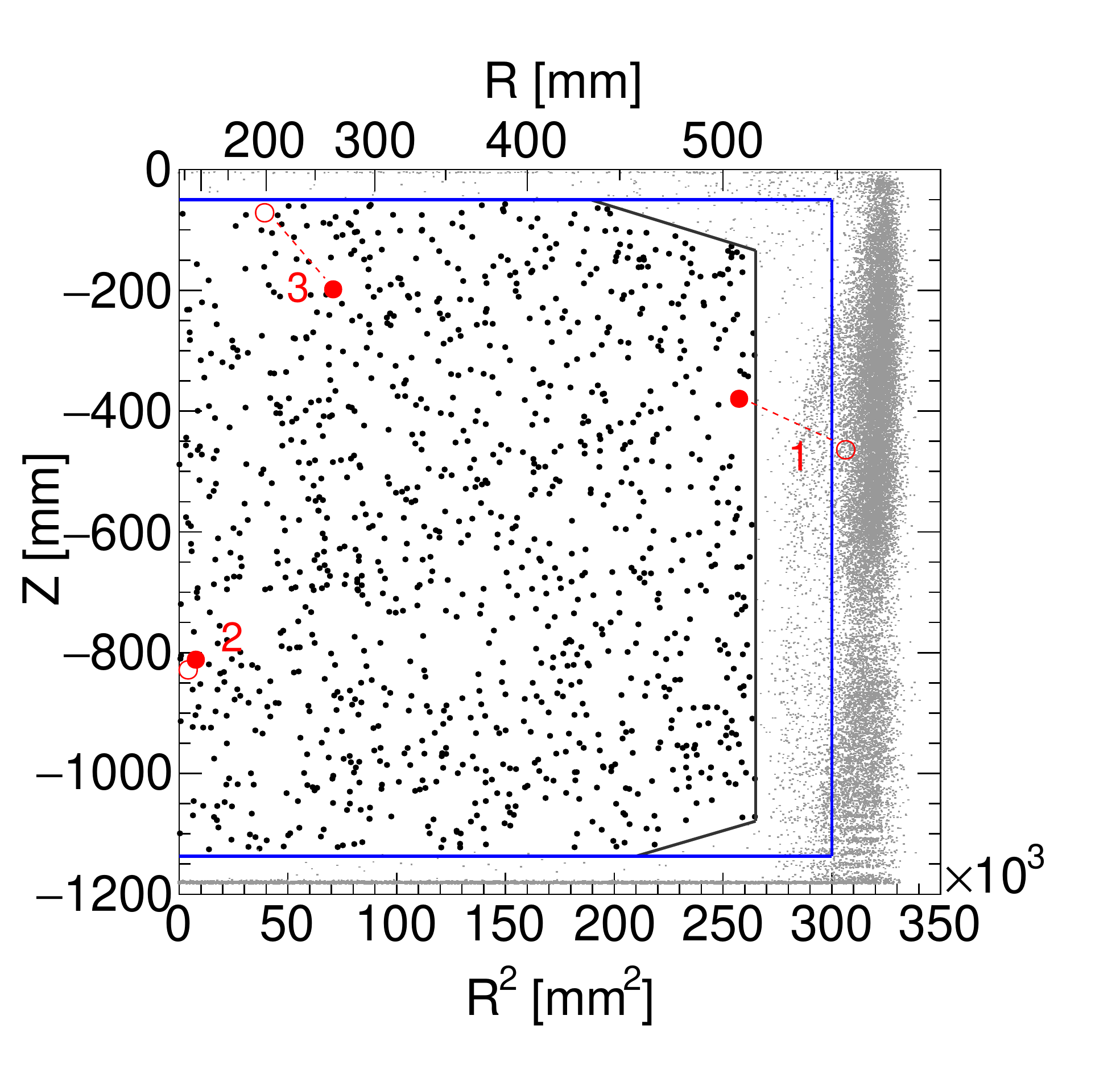}
  \caption{Left: MSNR candidates distribution on $\log_{10}(\sum{{\rm q}S2_{\rm b}}/\sum{{\rm q}S1})$ versus $\sum{{\rm q}S1}$.
  Three double scattering MSNR events (labeled red dots) left after MSNR 99\% upper acceptance cuts (red and magenta dashed lines) are applied.
  The dark matter search data \textcolor{black}{(single scattering events)} in data set~4 and data set~5 is also shown in black dots.
  		Right: MSNR candidates spatial distribution.
  The black and blue solid lines are the boundaries for FV~\cite{pandax4_first_paper} and LFV respectively.
  The solid circles represent the recoiling position which generate the largest $S2$.
  The hollow circles show the other recoiling position.
  The dashed lines connect the scatters that belong to the same event.
  Also, the single scattering events inside (outside) the FV are shown as black (grey) dots.
  }
  \label{fig:MSNR_candidate}
\end{figure}

In PandaX-4T commissioning data,
null MSNR candidate is found in data sets 1-2,
and 3 MSNR candidates are found in sets 3-5.
The statistical uncertainty is $\pm$~1.29~\cite{Feldman:1997qc} and $\pm$~1.73 for data sets 1-2 and data sets 3-5 respectively. 
The distribution of these three MSNR candidates 
on $\log_{10}(\sum{{\rm q}S2_{\rm b}}/\sum{{\rm q}S1})$ versus $\sum{{\rm q}S1}$ 
and the spatial distribution are shown in Fig.~\ref{fig:MSNR_candidate}. 
The $R_{\rm MC}$ (MSNR-to-SSNR) in different data sets are calculated based on MC simulation, 
and listed in Tab.~\ref{table:N_events_MC}
\textcolor{black}{(5.0 for pure neutron in data sets 1-2, 25.0 for neutron-X in data sets 1-2, 
5.0 for pure neutron in data sets 3-5, 20.8 for neutron-X in data sets 3-5
)}
.
Similar to MC method, the systematic uncertainty of $R_{\rm MC}$ comes from the simulation and the radioactivity measurement. 
For MSNR estimator, the simulation uncertainty is quantified by the difference between NR calibration data and NR MC simulation, 33\%. 
%The measurement uncertainty evaluation follows MC method, 21\%.
The $+1~\sigma$ upper limit of radioactivity gives around 21\% increment in $R_{\rm MC}$ (MSNR-to-SSNR).
In total, the systematic uncertainty of $R_{\rm MC}$ (MSNR-to-SSNR) is 40\%.
The predicted neutron-induced SSNR background events from MSNR data-driven method 
are 0.00~$\pm$~0.26 (pure neutron in data sets 1-2),
0.60~$\pm$~0.42 (pure neutron in data sets 3-5), 
0.00~$\pm$~0.05 (neutron-X in data sets 1-2),
0.14~$\pm$~0.10 (neutron-X in data sets 3-5) and summarized in Tab.~\ref{table:N_events_final}.

%MS candidates

\subsection{High energy gamma}
\label{chapter:NR_HEG_selection}
Compared with SSNR events and MSNR events, the number of HEG events is considerable in PandaX-4T detector, 
which indicates that HEG is a good estimator for SSNR background.
The HEGs from neutron capture can have energy of several MeV, much higher than the ER events produced by those radioactive isotopes in the detector materials. In addition, the HEG event may consist of many gamma particles and each gamma particle can scatter multiple times in the detector, which results in multiple $S1$s and $S2$s.
Specifically, the drift time of the HEG event is defined by the time separation of maximum $S1$ and maximum $S2$.
%The energy reconstruction formula follows that of SSNR, 
%however, considering the summation of all q$S1$ and the summation of all q$S2_{\rm b}$.
The energy reconstruction formula of SSNR is used for the HEG events with all summation of all q$S1$ and the summation of all q$S2_{\rm b}$.
Different from the low energy region, it is common that the PMTs get saturated when energy deposit reaches MeV scale or above.
To correct for the S2 after-pulsing effect and the PMT saturation effect,
an extra nonlinear correction factor is applied, 
which is derived from the difference between the reconstructed energy and true energy of the characteristic gamma rays in AmBe data, 
including 2.6~MeV ($^{208}$Tl), 4.4~MeV ($^{12}$C$^{*}$) and 9.3~MeV ($^{130}$Xe$^{*}$).

\begin{figure}[htbp]
\centering
	\includegraphics[width=0.45\columnwidth]{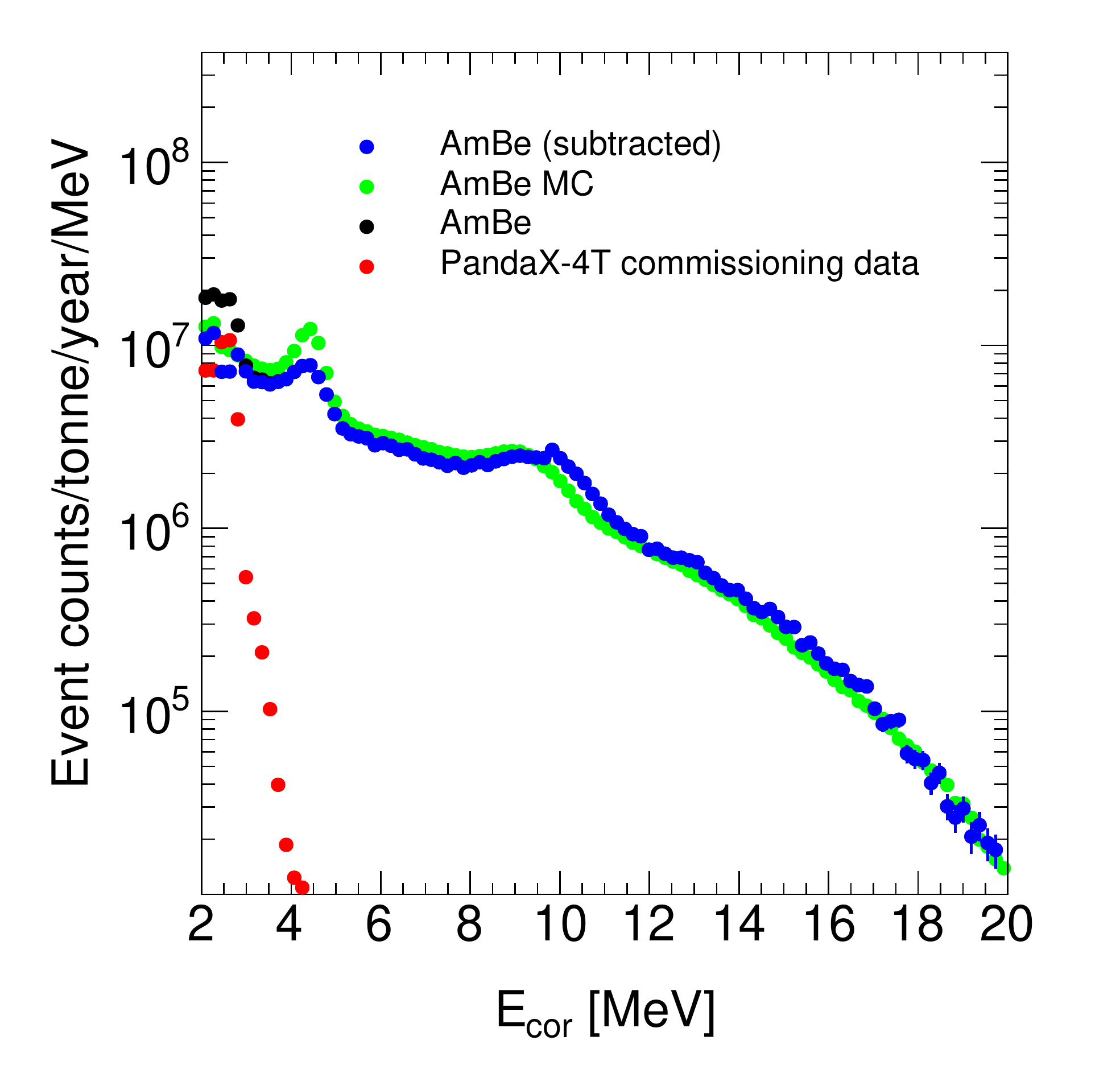}
	\includegraphics[width=0.45\columnwidth]{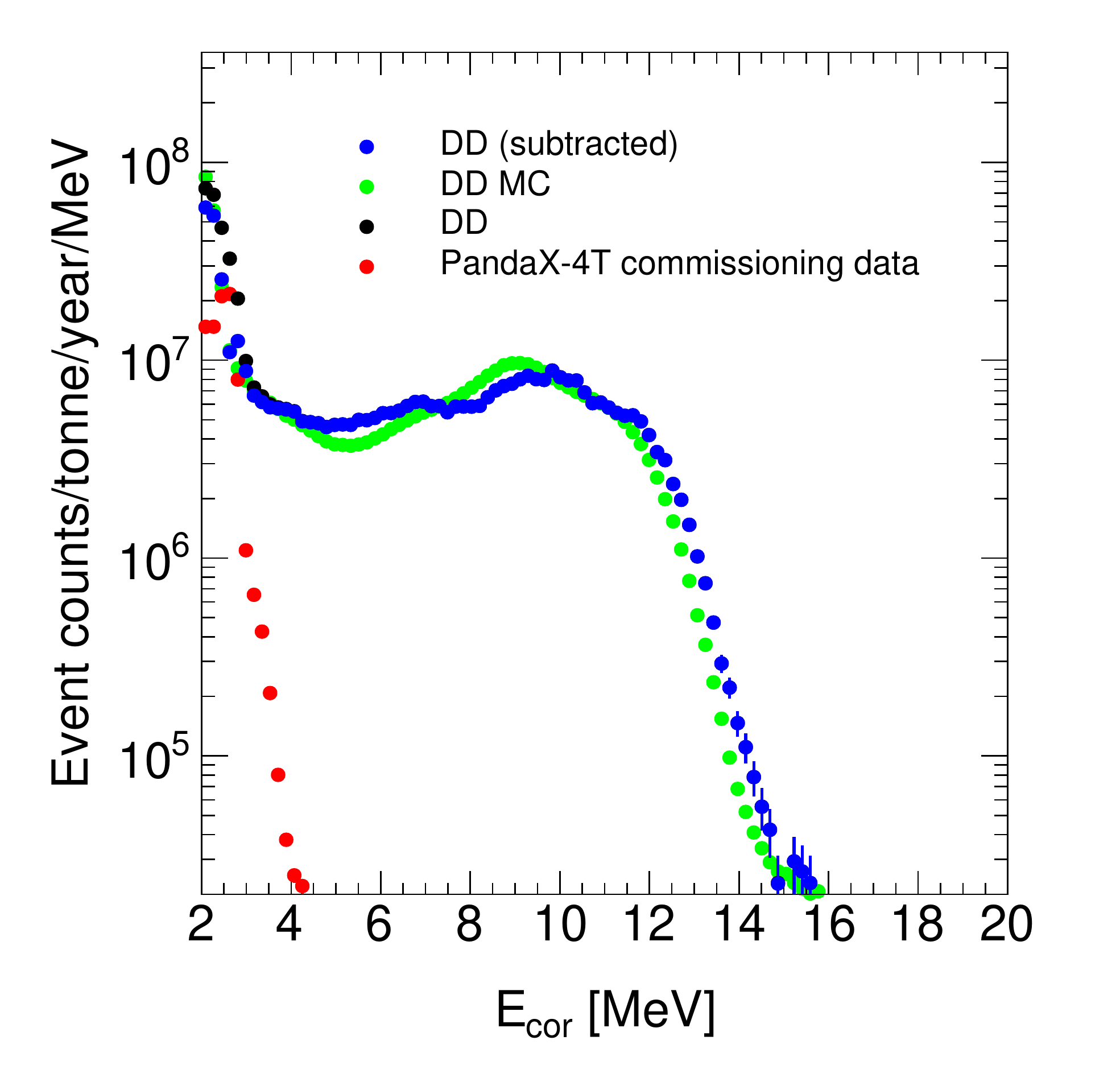}
    \caption{
    AmBe data (left), DD data (right) and MC high energy spectra comparison.
    The black dots represent the raw AmBe data or the raw DD data.
    The red dots represent the PandaX-4T commissioning data.
    The blue spectra are obtained by subtracting the PandaX-4T commissioning data from raw AmBe data or raw DD data.
    The MC simulation energy spectra are shown in green.
    }
\label{fig:HEG_Espectrum}
\end{figure}

To select the HEG candidates, the corrected energy $E_{\rm{cor}}$ should be limited to between 6~MeV and 20~MeV, 
which is similar to the previous work~\cite{p2neutron_paper}.
In the high energy region, multiple $S2$s sometimes overlap in the waveform.
The stray electrons, 
following the large $S2$ signals, can cause a long tail in the waveform. 
Also the PMT saturation may change the PMT charge distribution. 
Therefore, the reconstructed position is biased.
An extended fiducial volume (EFV) cut is adopted for HEG selection, 
the upper constraint on the position radius square $R^2$ is extended to $3.5 \times 10^5$~mm$^2$, 
the $z$ position is required to be 24~mm below the gate electrode and 100~mm above the cathode electrode.
The target mass of EFV is around 3.50 tonne.
Figure~\ref{fig:HEG_Espectrum} shows the energy spectra in EFV for the data and MC simulation.
The red, black, blue and green dots represent the PandaX-4T commissioning data, NR calibration data, 
NR calibration data with PandaX-4T commissioning data subtracted and the MC simulation, respectively. The overall shape of MC simulation agrees with the data.
The prediction of HEG candidates from MC simulation is validated through AmBe and DD calibration data, the MC simulation is consistent with data within 5\%, as shown in Tab.~\ref{table:MS_multiplicity}.

%HEG energy spectra

\begin{figure}[htbp]
\centering
\includegraphics[width=0.45\columnwidth]{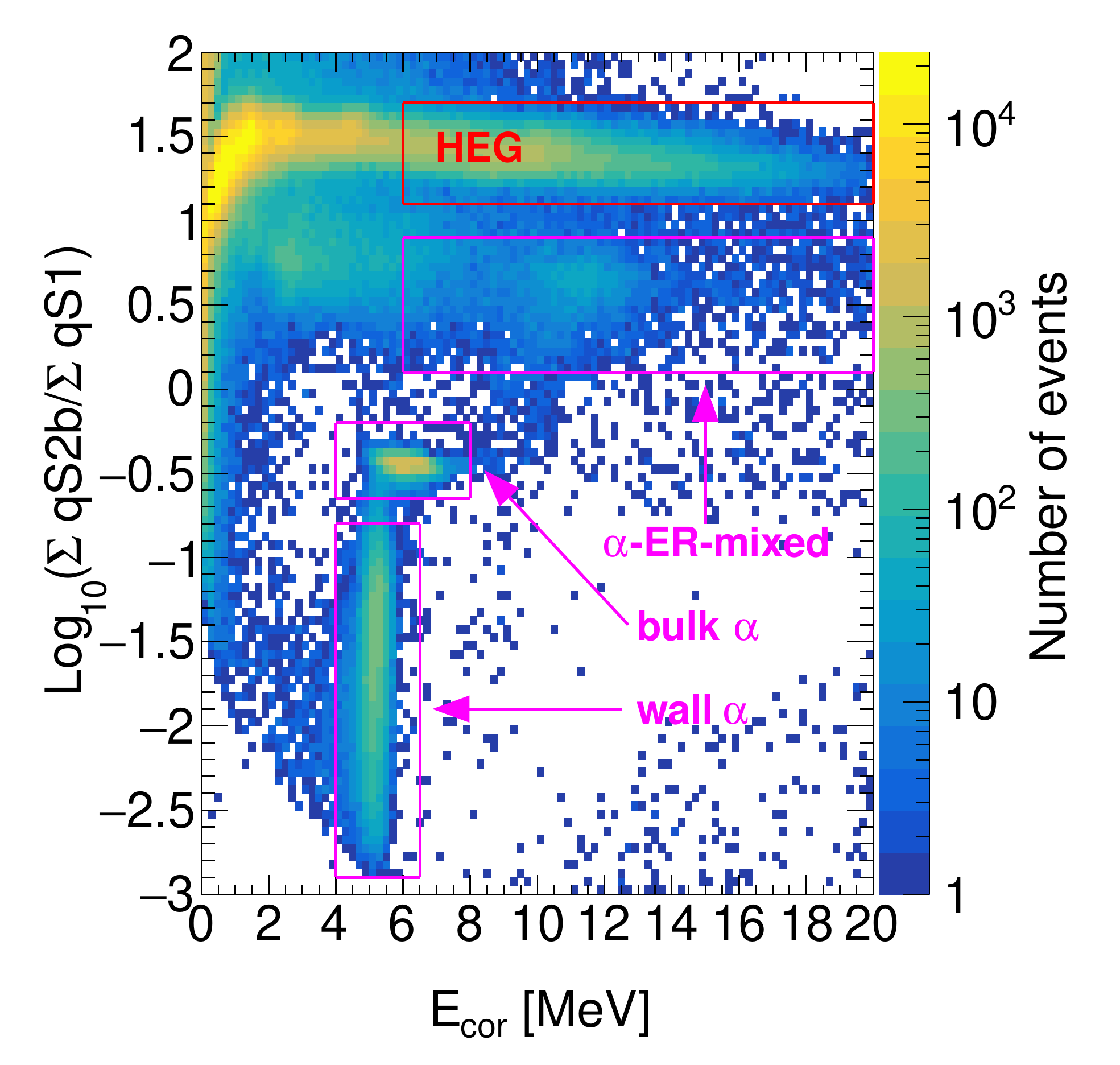}
\includegraphics[width=0.45\columnwidth]{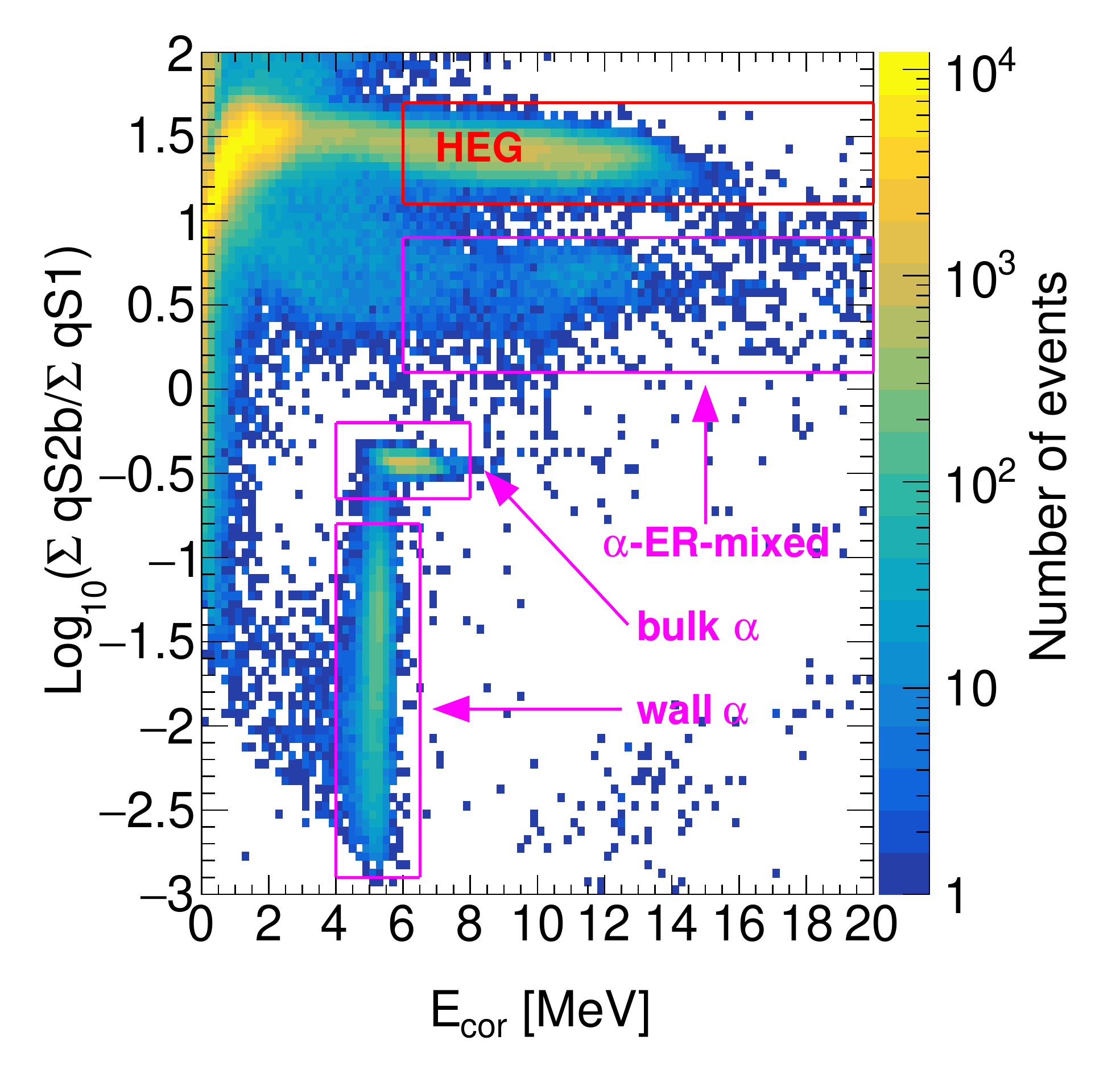}
\caption{The distribution of events in 
    $\log_{10}(\sum{{\rm q}S2_{\rm b}}/\sum{{\rm q}S1})$ versus $E_{\rm{cor}}$ in the AmBe data (left) and DD data (right).
    The HEG events located in the red box with y-axis within [1.1, 1.7] are distinguishable from other $\alpha$ events in magenta boxes. The $\alpha$-ER-mixed events have y-axis within [0.1, 0.9].}
\label{fig:HEG_S2S1_Ecor}
\end{figure}

Figure~\ref{fig:HEG_S2S1_Ecor} shows the distribution of $\log_{10}(\sum{{\rm q}S2_{\rm b}}/\sum{{\rm q}S1})$ versus $E_{\rm{cor}}$ from NR calibration data. The HEG candidates from neutron capture are located in the typical ER band with $\log_{10}(\sum{{\rm q}S2_{\rm b}}/\sum{{\rm q}S1})$ between 1.1 and 1.7. 
In the high energy region, there are also some $\alpha$ related events, labeled by bulk $\alpha$, wall $\alpha$ and $\alpha$-ER-mixed~\cite{p2neutron_paper}. Different from HEG, the $\alpha$-ER-mixed events
arise from the combination of $\alpha$ and gamma emissions (mainly with 2.6 MeV energy) from long lived radioisotopes in the detector materials.
The $\alpha$-ER-mixed events have $\log_{10}(\sum{{\rm q}S2_{\rm b}}/\sum{{\rm q}S1})$ mainly between 0.1 and 0.9,
and can have some leakage to the selection region of HEG candidates.

%HEG logS2S1 distribution

%HEG typical
% plotting code at P4-chain-dev/tmp
\begin{figure}[htbp]
\centering
\includegraphics[width=0.85\columnwidth]{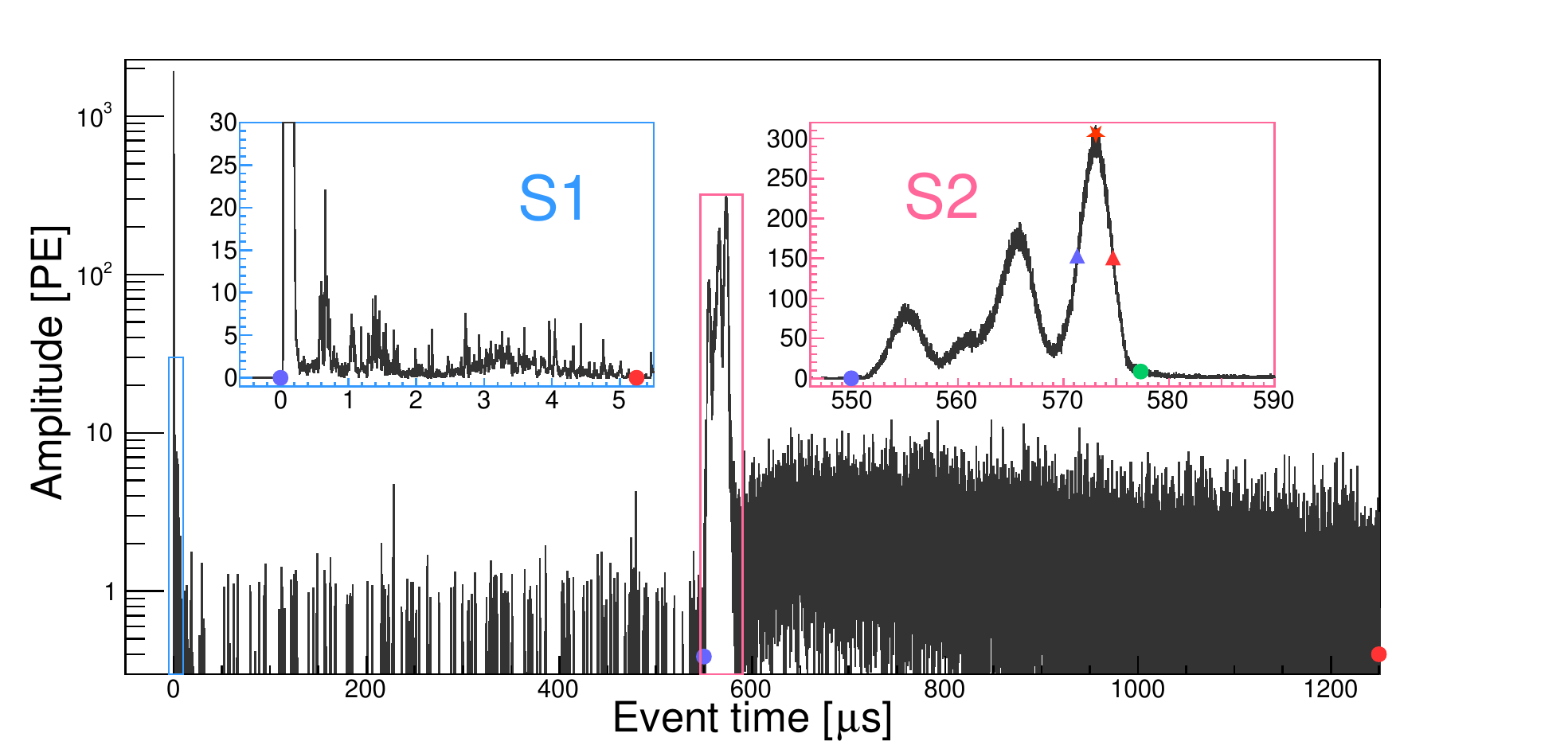}
\caption{Typical HEG event waveform. 
    The identified maximum $S1$ and identified maximum $S2$ are zoomed in the blue and pink insets, respectively.
    The violet dots and red dots indicate the start time and end time of each signal.
    Especially, they are not shown in the main graph for the maximum $S1$ in case of messy.
    For the maximum $S2$, the $w_{3\sigma}$ end boundary, the peak time, 
    the start boundary of the $w_{\rm fwhm}$ 
    and the end boundary of the $w_{\rm fwhm}$ are also plotted in the insets
    as the green dot, the red star, the violet triangle and the red triangle show, respectively.}
  \label{fig:HEG_waveform}
\end{figure}

A typical waveform of a HEG event is shown in Fig.~\ref{fig:HEG_waveform}. 
Some new waveform-based variables are defined and used for $\alpha$-ER-mixed event rejection later.
From these multiple $S2$ signals, the highest peak pulse is identified. 
The FWHM-width $w_{\rm fwhm}$ is defined as the full width at half maximum height of the highest peak pulse. 
Another width variable, $w_{3\sigma}$ is calculated 
from the signal start time to the equivalent +3~$\sigma$ Gaussian width of the highest peak pulse. 
Correspondingly, the charge can be obtained by integrating the relevant interval of the signal waveform. 
q$S2_{\rm b3\sigma}$ is the charge collected by the bottom PMT array 
by integrating over the $w_{3\sigma}$ interval of the $S2$ waveform. 
Generally, the stray electrons do not contribute to q$S2_{\rm b3\sigma}$.

To identify the suspicious $\alpha$-ER-mixed event in PandaX-4T commissioning data,
a boosted decision tree (BDT) technique is developed,
utilizing the TMVA (Toolkit for Multivariate Data Analysis) package in ROOT~\cite{TMVA}. The BDT inputs are based on the signal shape difference among the $\alpha$ candidates, the 2.6~MeV ER candidates and the HEG candidates, 
as listed in Tab.~\ref{table:BDT_variables}.

\begin{table*}[htbp]
\renewcommand\arraystretch{1.2}
\caption{The BDT input variables. }
\label{table:BDT_variables}
\doublerulesep 0.1pt \tabcolsep 10pt
\centering
\begin{tabular}{c|c|l}
\hline\hline
Variable & Unit & Explanation   \\ \hline
chargeRatio & 1 & The logarithm of the charge ratio to the base 10
($\log_{10}(\sum{\rm q}S2_{\rm b}/\sum{\rm q}S2_{{\rm b}3\sigma})$) \\
ratioTSignal & 1 & Time window percentage of all signals in the event waveform  \\
wS1 & sample (4ns) & Width of the largest $S1$   \\
qS1 & PE        & Raw charge of the largest $S1$   \\
q$S1$\underline{~}2nd & PE & Raw charge of the second largest $S1$ \\
widthTen$S1$ & sample (4ns) & Width of portion exceeding 10\%-height of the largest $S1$ \\
w$S1$CDF & sample (4ns) & Width of the largest $S1$ waveform enclosing 10\% to 90\% cumulative charge \\
h$S1$  & PE/sample   & Height of the largest $S1$   \\
q$S2$  & PE          & Raw charge of the largest $S2$   \\
w$S2$  & sample (4ns) & Width of the largest $S2$    \\
widthTen$S2$ & sample (4ns) & Width of portion exceeding 10\%-height of the highest peak in the largest $S2$ \\
w$S2$CDF & sample (4ns) & Width of the largest $S2$ waveform enclosing 10\% to 90\% cumulative charge \\
h$S2$  & PE/sample   & Height of the largest $S2$   \\
$S1$Asy & 1          & Ratio of top and bottom charge difference over the total charge for the maximum $S1$ \\
$S2$Asy & 1          & Ratio of top and bottom charge difference over the total charge for the maximum $S2$ \\
\hline\hline
\end{tabular}
\end{table*}

%P4 HEG canditates
\begin{figure}[htbp]
  \centering
  \includegraphics[width=0.45\columnwidth]{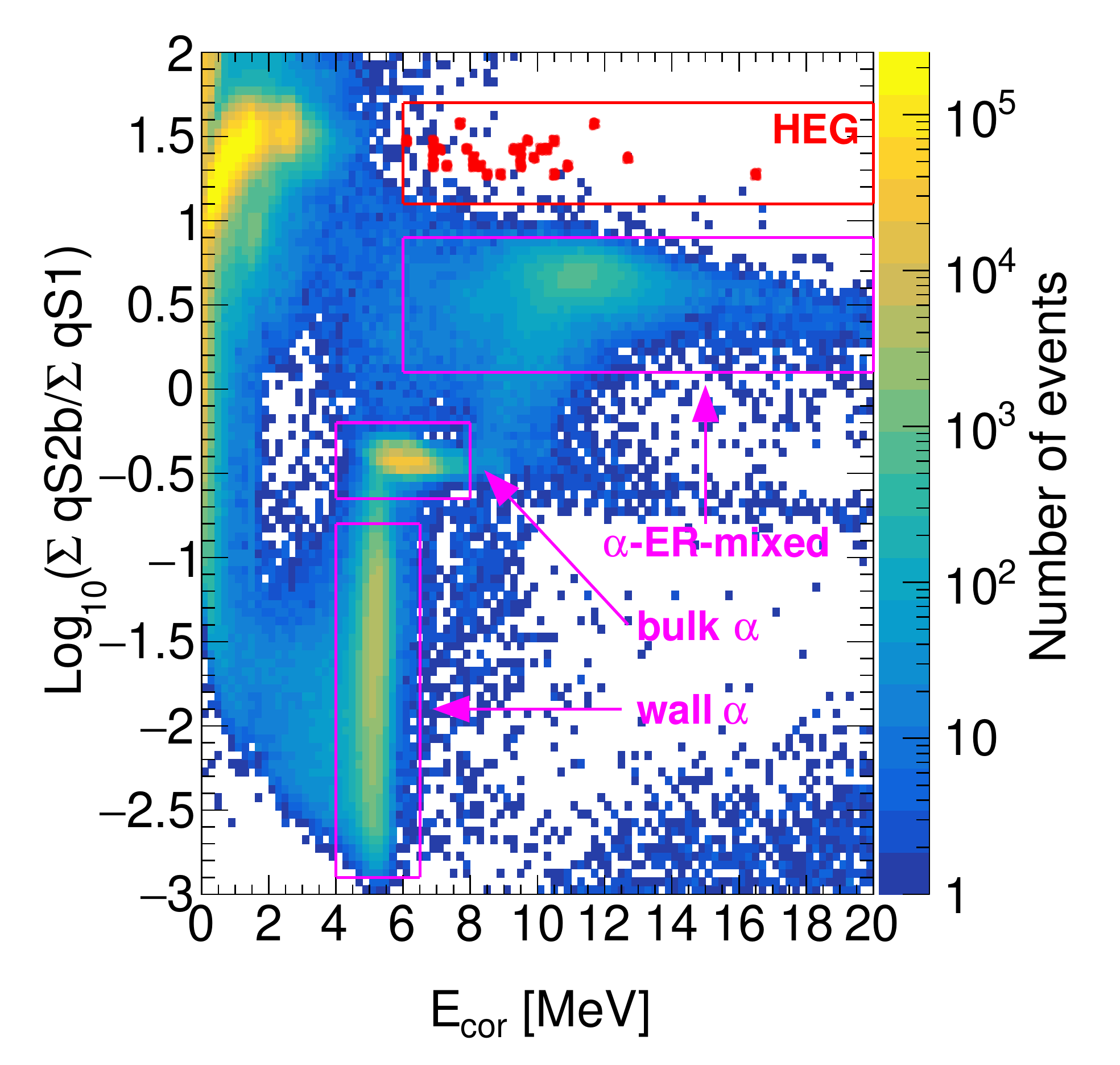}
  \includegraphics[width=0.45\columnwidth]{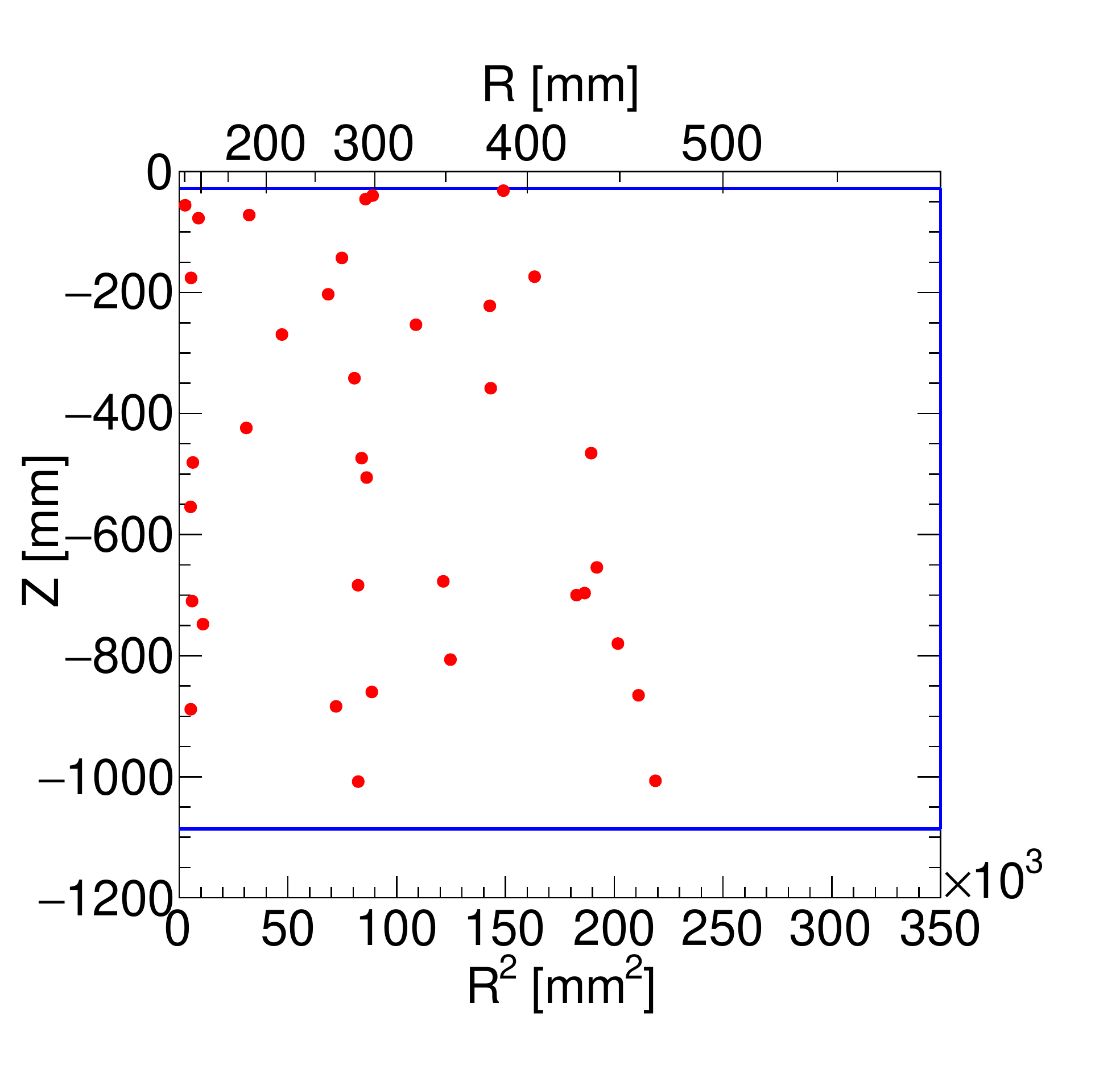}
  \caption{Left: Event distribution on 
  $\log_{10}(\sum{{\rm q}S2_{\rm b}}/\sum{{\rm q}S1})$ versus $E_{\rm{cor}}$ in all PandaX-4T commissioning data
  before BDT cut is applied.
  The red box indicates the HEG event selection region 
  and the magenta boxes represent the $\alpha$ relevant event selection regions.
  \textcolor{black}{
  The HEG events after BDT cut are highlighted in red dots.
  Right: The position distributions for the HEG events.
  Only the positions that are calculated from the largest $S1$ and largest $S2$ are shown.
  The blue solid lines indicate the EFV boundaries.}
  }
  \label{fig:P4_HEG}
\end{figure}

The events, inside the HEG selection region (top red box in Fig.~\ref{fig:HEG_S2S1_Ecor}) in 21.6~hours AmBe data,
are used for HEG training.
Also, the events within the $\alpha$-ER-mixed event region (top magenta box in \textcolor{black}{Fig.~\ref{fig:P4_HEG}}) in the 30 days data in PandaX-4T commissioning data set~4 are used for $\alpha$-ER-mixed event training.
The remaining NR calibration data and the other PandaX-4T commissioning data are selected to derive the HEG selection efficiency and the $\alpha$-ER-mixed event rejection power, respectively.
The BDT variables distribution, their correlation matrices
and the BDT performance 
are shown in the appendix (Fig.~\ref{fig:BDT_variables}, Fig.~\ref{fig:BDT_correlationn_matrix} and Fig.~\ref{fig:BDT_performance}).
\textcolor{black}{
In the BDT model training process, a strict $S2$Asy cut is applied as Fig.~\ref{fig:BDT_variables} shows.
It ensures the purity of HEG training samples.
However, it is not applied to the testing samples.
Also, it can be identified that $\alpha$-ER-mixed events have two components.
They are from the $\alpha$ decay of $^{222}$Rn and $^{218}$Po, 
which release $\alpha$ particles with the energy of 5.5~\si{\MeV} and 6~\si{\MeV} respectively.
}
The most powerful variables are q$S2$, w$S2$, h$S1$ and q$S1$\underline{~}2nd.
%The BDT cut is set at 0.2 and the HEG event selection efficiency and the $\alpha$-ER-mixed event rejection power are both 100\%, 
Due to the significant difference in signal shape and $S2$ charge,
over 99.9\% of $\alpha$-ER-mixed events are rejected while keeping an almost 100\% efficiency for HEG events.

The PandaX-4T commissioning data event distribution 
on $\log_{10}(\sum{{\rm q}S2_{\rm b}}/\sum{{\rm q}S1})$ versus $E_{{\rm cor}}$
is plotted in Fig.~\ref{fig:P4_HEG}.
Note that the electric field condition of the first two data sets in PandaX-4T commissioning data
is different from that of other data sets.
The corresponding charges of $S2$s in these two data sets are scaled for consistency, 
depending on the SEG$\times$EEE ratio in Ref~\cite{pandax4_first_paper}.
In total, 102 high energy events are inside the HEG selection region.
After the BDT cut applied, only 36 HEG candidates survive, 
where 4 candidates are from data sets 1-2 and 32 candidates are from data sets 3-5.
\textcolor{black}{
The reconstructed position distribution of these HEG events are also shown in Fig.~\ref{fig:P4_HEG}.
Due to the multiple-pulses S2 waveform and saturation effect, 
the reconstructed positions in $R^2$ direction are biased.
}
The statistical uncertainty for HEG method is $\pm~2$ and $\pm$~5.66 for data sets 1-2 and data sets 3-5 respectively.
The $R_{\rm MC}$ (HEG-to-SSNR) in different data sets are calculated based on MC simulation, 
and listed in Tab.~\ref{table:N_events_MC}
\textcolor{black}{(32.1 for pure neutron in data sets 1-2, 160.2 for neutron-X in data sets 1-2, 
30.3 for pure neutron in data sets 3-5, 126.1 for neutron-X in data sets 3-5
)}
.
The systematic uncertainty of $R_{\rm MC}$ (HEG-to-SSNR) comes from radioactivity measurement and simulation as well.
Following MSNR method, the measurement uncertainty of $R_{\rm MC}$ (HEG-to-SSNR) is 19\%.
The simulation uncertainty of $R_{\rm MC}$ (HEG-to-SSNR) is only 5\%, as Tab.~\ref{table:MS_multiplicity} shows.
The total systematic uncertainty of $R_{\rm MC}$ (HEG-to-SSNR) is 20\%.
The predicted neutron-induced SSNR background events from HEG data-driven method 
are 0.12~$\pm$~0.076 (pure neutron in data sets 1-2),
1.06~$\pm$~0.28 (pure neutron in data sets 3-5), 
0.03~$\pm$~0.01 (neutron-X in data sets 1-2),
0.25~$\pm$~0.07 (neutron-X in data sets 3-5) and summarized in Tab.~\ref{table:N_events_final}.

\begin{table*}[htbp]
\renewcommand\arraystretch{1.2}
\footnotesize
\caption{Predicted neutron background events in PandaX-4T commissioning data in the unit of counts.
In a conservative way, 
the uncertainties in the ``Total'' column is the summation of that in each data sets,
as they are not completely independent.
Also, the final uncertainties in the ``Average'' column are assumed to be 50\%.
}
\doublerulesep 0.1pt \tabcolsep 10pt
\centering
\begin{tabular}{cc|cc|c|c}
\hline\hline
\multicolumn{2}{c|}{Data sets}                            & Set 1-2   & Set 3-5   & Total     & Average         \\ \hline
\multicolumn{1}{c|}{\multirow{3}{*}{Pure neutron}} & MC method  & 0.12~$\pm$~0.06 & \textcolor{black}{0.62~$\pm$~0.32} & \textcolor{black}{0.74~$\pm$~0.38} & \multirow{3}{*}{0.93~$\pm$~0.46} \\
\multicolumn{1}{c|}{}                              & MSNR method & 0~$\pm$~0.26 & 0.60~$\pm$~0.42 & 0.60~$\pm$~0.68 &  \\
\multicolumn{1}{c|}{}                              & HEG method & 0.12~$\pm$~0.07 & 1.06~$\pm$~0.28 & 1.18~$\pm$~0.35 &   \\ \hline
\multicolumn{1}{c|}{\multirow{3}{*}{Neutron-X}}    & MC method & 0.02~$\pm$~0.01 & \textcolor{black}{0.15~$\pm$~0.08} & \textcolor{black}{0.17~$\pm$~0.09} & \multirow{3}{*}{0.22~$\pm$~0.11} \\
\multicolumn{1}{c|}{}                              & MSNR method & 0~$\pm$~0.05 & 0.14~$\pm$~0.10 & 0.14~$\pm$~0.15 & \\
\multicolumn{1}{c|}{}                              & HEG method & 0.03~$\pm$~0.01 & 0.25~$\pm$~0.07 & 0.28~$\pm$~0.08 &  \\
\hline\hline
\end{tabular}
\label{table:N_events_final}
\end{table*}

\section{Summary}
\label{chapter:N_estimation}

%final estimation number
In summary, a robust neutron background estimation for PandaX-4T commissioning data is implemented,
based on the MC simulation and the neutron relevant feature events.
The MC simulation method is conventional and straightforward.
The data-driven methods further make use of the feature events, MSNR and HEG, in the data.
These data-driven results and MC estimations agree within uncertainties.
The weighted averages of them give the neutron backgrounds,
i.e., $0.93 \pm 0.46$ pure neutron and $0.22 \pm 0.11$ neutron-X.
%Meanwhile, a conservative 50\% uncertainty of the background is adopted in Ref.~\cite{pandax4_first_paper}.
Compared with previous work~\cite{pandax4t_sensitivity_paper,p2neutron_paper}, 
this analysis gives more credible results. 
Considering other backgrounds~\cite{pandax4_first_paper},
this result shows that the neutron background is subdominant and well controlled in PandaX-4T experiment.

\acknowledgements
This project is supported in part 
by a grant from the Ministry of Science and Technology of China (No.~2016YFA0400301),
grants from National Science Foundation of China 
(Nos.~12090060, 12005131m 11905128, 11925502, 11775141),
and by Office of Science and Technology,
Shanghai Municipal Government (grant No.~18JC1410200).
We thank supports from Double First Class Plan of the Shanghai Jiao Tong University.
We also thank the sponsorship 
from the Chinese Academy of Science Center for Excellence in Particle Physics (CCEPP),
Hongwen Foundation in Hong Kong, China, and Tencent Foundation in China.
Finally, we thank the CJPL administration and the Yalong River Hydropower Devepolment Company Ltd. for indispensable logistical support and other help.

\bibliographystyle{unsrt}
\bibliography{reference}

\begin{thebibliography}{10}

\bibitem{Plank_newest_result}
Nabila Aghanim et~al.
\newblock {Planck 2018 results-VI. Cosmological parameters}.
\newblock {\em Astronomy \& Astrophysics}, 641:A6, 2020.

\bibitem{jianglai_xun_xiangdong_review}
Jianglai Liu et~al.
\newblock {Current status of direct dark matter detection experiments}.
\newblock {\em Nature Physics}, 13(3):212--216, 2017.

\bibitem{zhaolireview}
Li~Zhao and Jianglai Liu.
\newblock {Experimental search for dark matter in China}.
\newblock {\em Front. Phys. (Beijing)}, 15(4):44301, 2020.

\bibitem{CJPL_intro}
Yu-Cheng Wu et~al.
\newblock {Measurement of Cosmic Ray Flux in China JinPing underground
  Laboratory}.
\newblock {\em Chin. Phys. C}, 37(8):086001, 2013.

\bibitem{CJPL_intro_JNE}
Ziyi Guo et~al.
\newblock {Muon flux measurement at China Jinping Underground Laboratory}.
\newblock {\em Chin. Phys. C}, 45(2):025001, 2021.

\bibitem{CJPL2_intro}
Zhao-Ming Zeng, Hui Gong, Jian-Min Li, Qian Yue, Zhi Zeng, and Jian-Ping Cheng.
\newblock {Design of the thermal neutron detection system for CJPL-II}.
\newblock {\em Chin. Phys. C}, 41(5):056002, 2017.

\bibitem{finalcpc}
Qiuhong Wang et~al.
\newblock {Results of dark matter search using the full PandaX-II exposure}.
\newblock {\em Chin. Phys. C}, 44(12):125001, 2020.

\bibitem{Xia:2018qgsPandaXII}
Jingkai Xia et~al.
\newblock {PandaX-II Constraints on Spin-Dependent WIMP-Nucleon Effective
  Interactions}.
\newblock {\em Phys. Lett. B}, 792:193--198, 2019.

\bibitem{PandaX-II:2020udv}
Xiaopeng Zhou et~al.
\newblock {A search for solar axions and anomalous neutrino magnetic moment
  with the complete PandaX-II data}.
\newblock {\em Chin. Phys. Lett.}, 38(1):011301, 2021.

\bibitem{PandaX-II:2021nsg}
Chen Cheng et~al.
\newblock {Search for Light Dark Matter-Electron Scatterings in the PandaX-II
  Experiment}.
\newblock {\em Phys. Rev. Lett.}, 126(21):211803, 2021.

\bibitem{pandax4_first_paper}
Yue Meng et~al.
\newblock {Dark Matter Search Results from the PandaX-4T Commissioning Run}.
\newblock {\em Phys. Rev. Lett.}, 127:261802, Dec 2021.

\bibitem{pandax4t_sensitivity_paper}
Hongguang Zhang et~al.
\newblock {Dark matter direct search sensitivity of the PandaX-4T experiment}.
\newblock {\em Sci. China Phys. Mech. Astron.}, 62(3):31011, 2019.

\bibitem{alpha_n_F19}
EB~Norman et~al.
\newblock {19F ($\alpha$, n) thick target yield from 3.5 to 10.0 MeV}.
\newblock {\em Applied Radiation and Isotopes}, 103:177--178, 2015.

\bibitem{alpha_n_production}
V.~A. Kudryavtsev, P.~Zakhary, and B.~Easeman.
\newblock {Neutron production in ($\alpha, n$) reactions}.
\newblock {\em Nucl. Instrum. Meth. A}, 972:164095, 2020.

\bibitem{sources4a_code}
W~Betal Wilson et~al.
\newblock {Sources: a code for calculating ($\alpha$, n), spontaneous fission,
  and delayed neutron sources and spectra}.
\newblock {\em Radiation protection dosimetry}, 115(1-4):117--121, 2005.

\bibitem{p2neutron_paper}
Qiuhong Wang et~al.
\newblock {An Improved Evaluation of the Neutron Background in the PandaX-II
  Experiment}.
\newblock {\em Sci. China Phys. Mech. Astron.}, 63(3):231011, 2020.

\bibitem{pandax2calibration}
Binbin Yan et~al.
\newblock {Determination of responses of liquid xenon to low energy electron
  and nuclear recoils using a PandaX-II detector}.
\newblock {\em Chin. Phys. C}, 45(7):075001, 2021.

\bibitem{pandax4t_cryo_design}
Li~Zhao et~al.
\newblock {The cryogenics and xenon handling system for the PandaX-4T
  experiment}.
\newblock {\em JINST}, 16(06):T06007, 2021.

\bibitem{pandax_SS_vessel}
Tao Zhang et~al.
\newblock {Low Background Stainless Steel for the Pressure Vessel in the
  PandaX-II Dark Matter Experiment}.
\newblock {\em JINST}, 11(09):T09004, 2016.

\bibitem{LUX_DDcalibration}
D.~S. Akerib et~al.
\newblock {Low-energy (0.7-74 keV) nuclear recoil calibration of the LUX dark
  matter experiment using D-D neutron scattering kinematics}.
\newblock 8 2016.

\bibitem{DDgenerator_paper}
De-Dong He et~al.
\newblock Design and optimization of thermal neutron device based on
  deuterium-deuterium neutron generator.
\newblock {\em Fusion Engineering and Design}, 166:112289, 2021.

\bibitem{csikai1987}
Julius Csikai.
\newblock {CRC handbook of fast neutron generators}.
\newblock 1987.

\bibitem{qzcmaterial}
Zhicheng Qian et~al.
\newblock {Low Radioactive Material Screening and Background Control for the
  PandaX-4T Experiment}.
\newblock 12 2021.

\bibitem{geant4_collaboration}
Sea Agostinelli et~al.
\newblock Geant4—a simulation toolkit.
\newblock {\em Nuclear instruments and methods in physics research section A:
  Accelerators, Spectrometers, Detectors and Associated Equipment},
  506(3):250--303, 2003.

\bibitem{xenon1t_analysis_paper2}
E.~Aprile et~al.
\newblock {XENON1T dark matter data analysis: Signal and background models and
  statistical inference}.
\newblock {\em Phys. Rev. D}, 99(11):112009, 2019.

\bibitem{neutron_bkg_estimation_2004}
M.J. Carson et~al.
\newblock Neutron background in large-scale xenon detectors for dark matter
  searches.
\newblock {\em Astroparticle Physics}, 21(6):667--687, 2004.

\bibitem{neutron_mean_path}
Peter Forck.
\newblock {Machine and People Protection}.
\newblock In {\em {CAS - CERN Accelerator School 2019}: {Introduction to
  Accelerator Physics}}, 5 2021.

\bibitem{p2_pos_rec_paper}
Dan Zhang et~al.
\newblock {Horizontal position reconstruction in PandaX-II}.
\newblock {\em JINST}, 16(11):P11040, 2021.

\bibitem{Feldman:1997qc}
Gary~J. Feldman and Robert~D. Cousins.
\newblock {A Unified approach to the classical statistical analysis of small
  signals}.
\newblock {\em Phys. Rev. D}, 57:3873--3889, 1998.

\bibitem{TMVA}
Andreas Hocker et~al.
\newblock {TMVA - Toolkit for Multivariate Data Analysis}.
\newblock 3 2007.

\end{thebibliography}

\begin{appendix}

\renewcommand{\thesection}{Appendix}

\section{}

%BDT variables
\begin{figure}
  \centering
  \includegraphics[width=0.85\columnwidth]{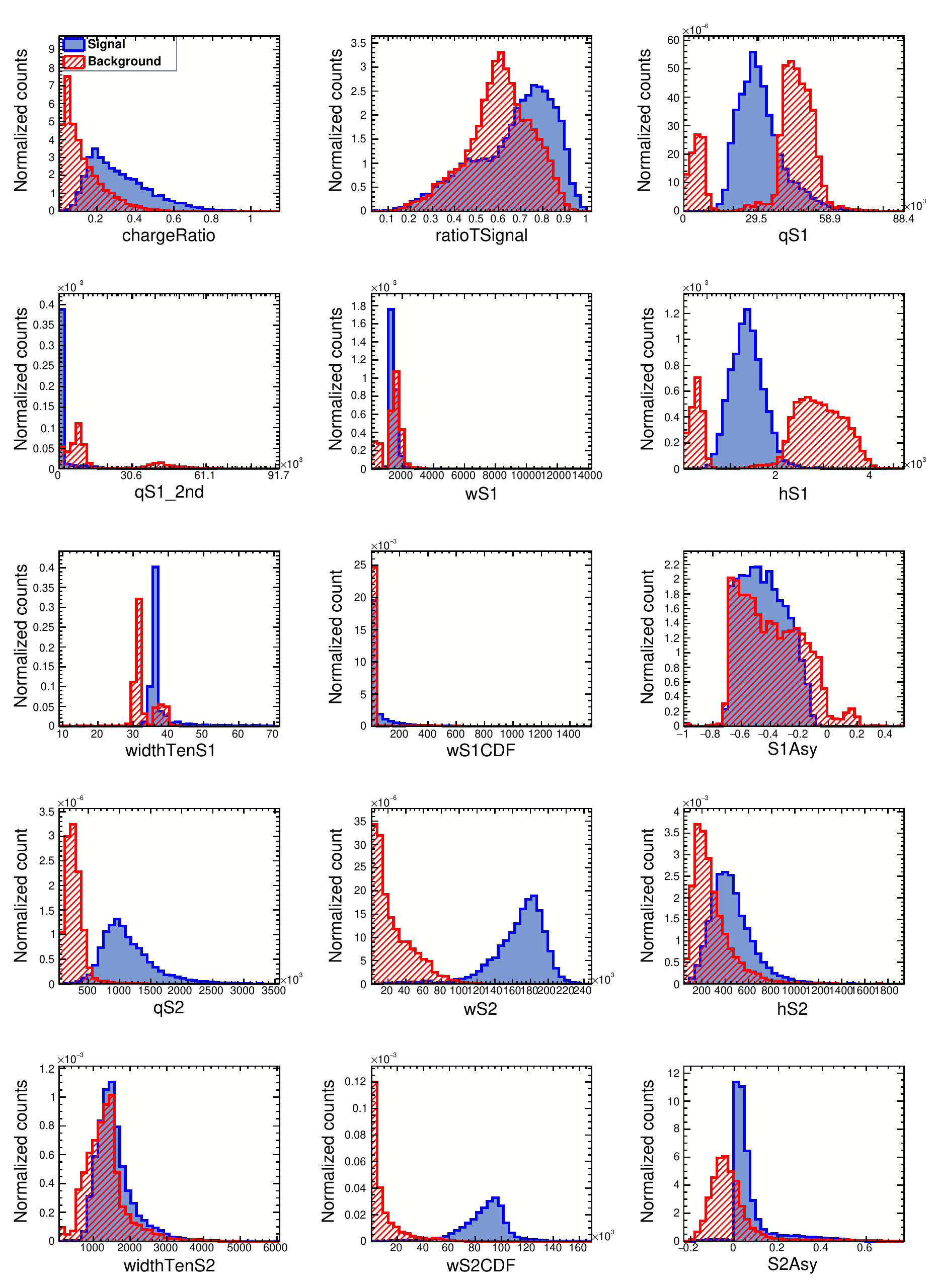}
  \caption{Distribution of the selected variables from the HEG event (signal) and the $\alpha$-ER-mixed event (background). The HEG event distribution is shown in solid blue and the $\alpha$-ER-mixed event distribution in hatched red.}
  \label{fig:BDT_variables}
\end{figure}

%BDT correlation matrix
\begin{figure}
  \centering
  \includegraphics[width=0.45\columnwidth]{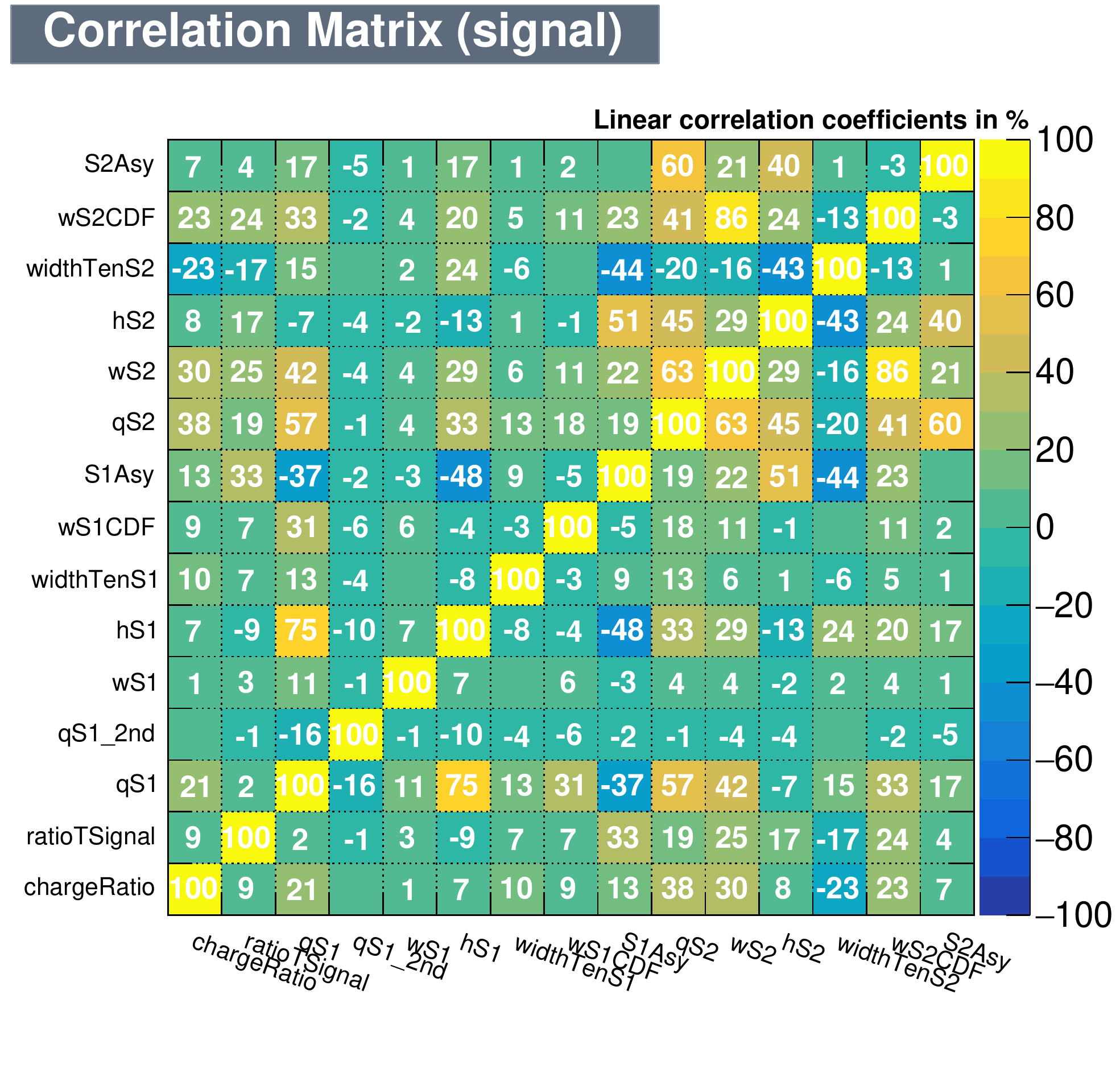}
  \includegraphics[width=0.45\columnwidth]{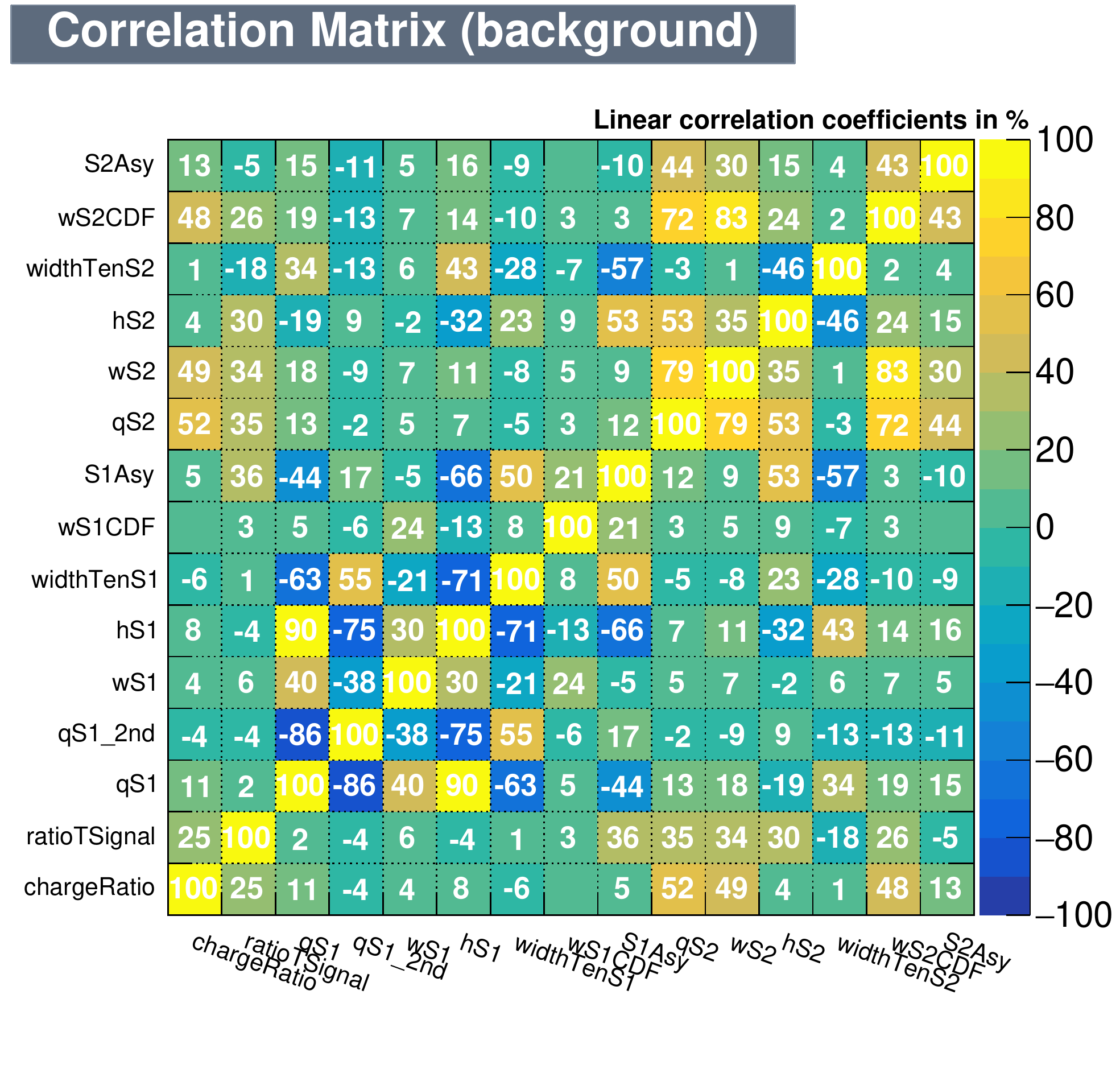}
  \caption{Correlation matrices for signal and background samples.}
  \label{fig:BDT_correlationn_matrix}
\end{figure}

%BDT
\begin{figure}
  \centering
  \includegraphics[width=0.45\columnwidth]{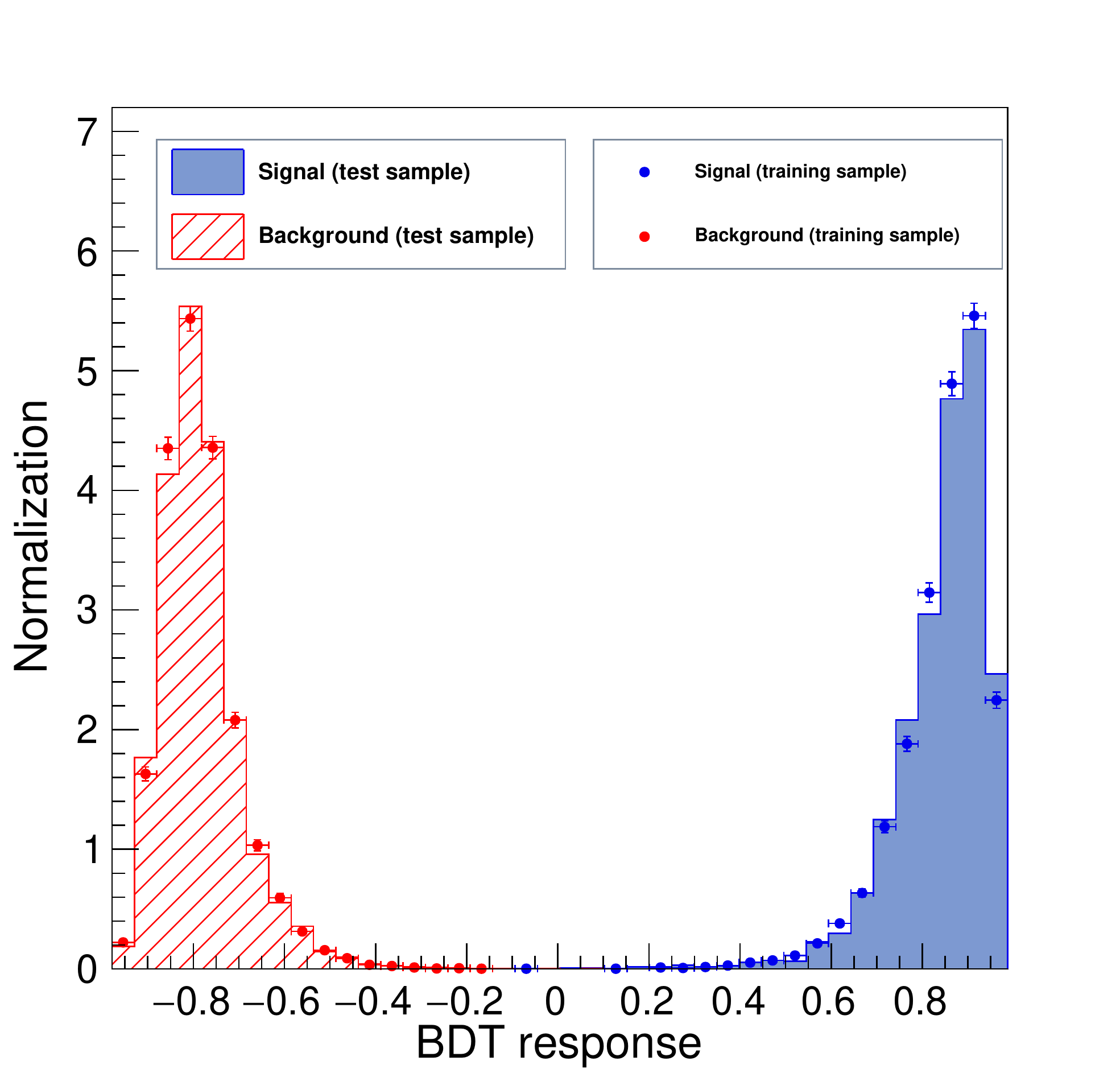}
  \includegraphics[width=0.45\columnwidth]{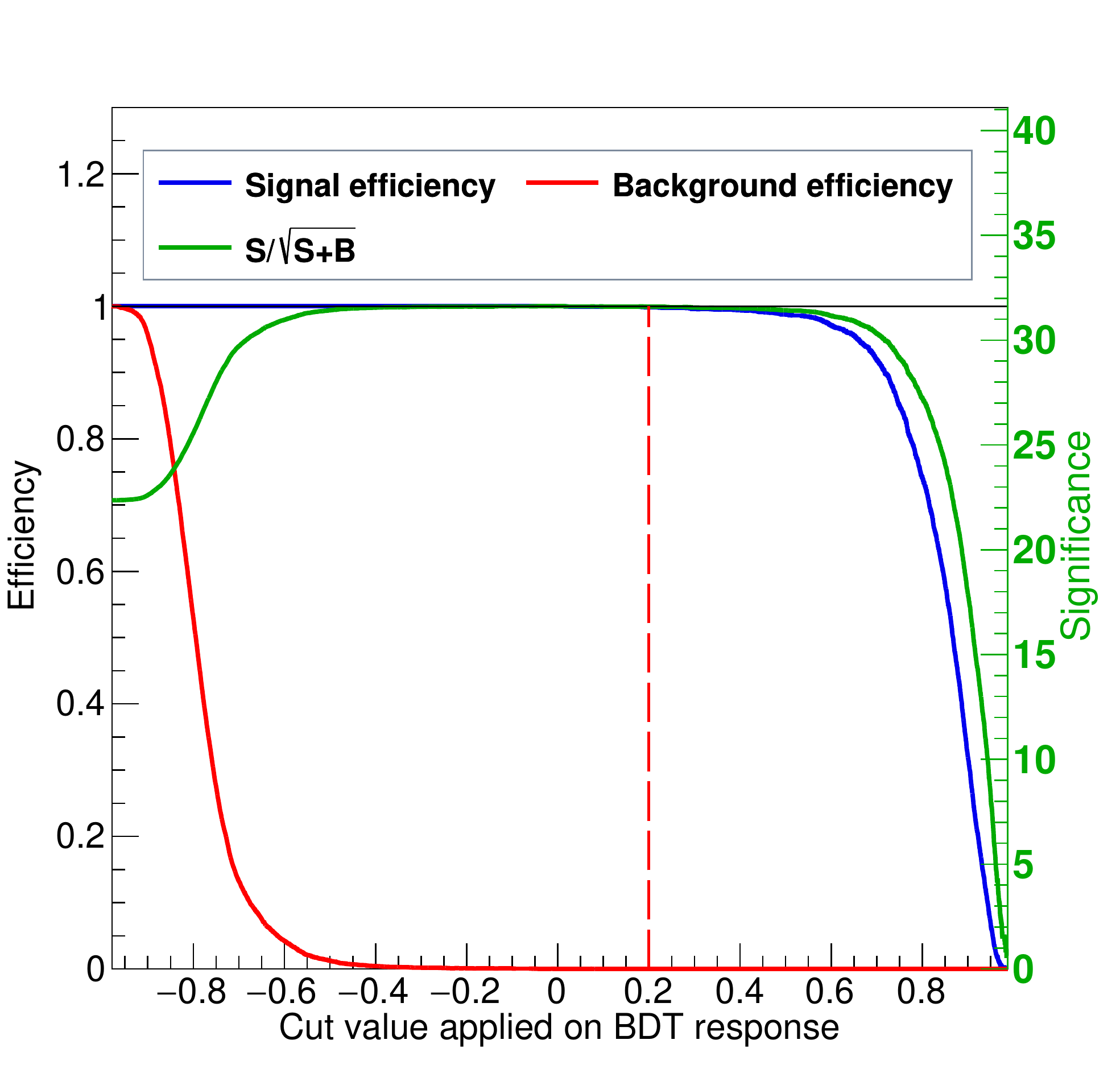}
  \caption{Left: Distribution of TMVA discriminator for HEG and $\alpha$-ER-mixed event sample.
  	Right: Signal efficiency, background efficiency, 
  	and discrimination significance $(S/\sqrt{S+B})$ versus the BDT response cut values. 
  	\textcolor{black}{The final BDT cut value is overlaid as the red dashed line shows.}}
  \label{fig:BDT_performance}
\end{figure}

\end{appendix}

\end{document}